\documentclass[12pt, a4paper]{article}
\usepackage{authblk}
\usepackage{hyperref}
\usepackage{soul}
\usepackage{longtable}
\usepackage{comment}
\usepackage{multirow}
\usepackage{lineno}
\usepackage{setspace}

\usepackage[utf8]{inputenc}
\usepackage[ngerman, english]{babel} 
\usepackage{eurosym} 
\usepackage{amsmath}
\usepackage{tabularx} 
\usepackage{threeparttable} 
\usepackage{multirow} 
\usepackage{booktabs} 
\usepackage{pdflscape}
\usepackage{lscape} 
\usepackage{longtable}
\usepackage{dcolumn}
\usepackage[rounding]{rccol} 
\usepackage{morefloats} 
\usepackage{array}
\usepackage{fltpoint}[2004/11/12]
\usepackage{dcolumn}
\usepackage{rccol}
\usepackage{numprint}
\usepackage{listings}
\usepackage{csquotes}
\usepackage{lscape}
\usepackage{longtable} 
\usepackage[T1]{fontenc}
\usepackage{adjustbox}
\usepackage{numprint}

\usepackage[font=small]{caption}
\usepackage{subcaption}
\usepackage{dsfont}

\usepackage{listings} 
\usepackage[authoryear]{natbib}

\usepackage{amsthm}
\usepackage{amsmath}
\usepackage{amsbsy}
\usepackage{amssymb}
\usepackage{comment} 

\usepackage[left=2.5cm, top=2.5cm, bottom=2.5cm, right=2.5cm]{geometry} 
\usepackage{setspace} 
\usepackage{tocloft} 
\usepackage{hyperref}
\usepackage{yhmath}
\usepackage{rotating}
\usepackage{graphicx}

\usepackage{mathtools}

\usepackage{tabularx}
\usepackage{tikz}
\usetikzlibrary{arrows}

\usetikzlibrary{decorations,arrows}
\usetikzlibrary{decorations.pathmorphing}
\usepgflibrary{decorations.pathreplacing} 
\usetikzlibrary{decorations.fractals}
\usetikzlibrary{decorations.footprints}
\usetikzlibrary{decorations.shapes}
\usetikzlibrary{decorations.text}
\usetikzlibrary{decorations.pathmorphing}
\usetikzlibrary{patterns}
\usepackage{epsfig}
 \usepackage{pst-grad} 
 \usepackage{pst-plot} 
\usepackage{filecontents}

\usepackage[font=small, labelfont=bf, justification=centering, singlelinecheck=false]{caption}

\renewcommand\appendix{\par%
\setcounter{section}{0}%
\setcounter{figure}{0}%
\setcounter{table}{0}%
\renewcommand\thesection{\Alph{section}}%
\renewcommand\thetable{\Alph{section}.\arabic{table}}%
\renewcommand\thefigure{\Alph{section}.\arabic{figure}}} %

\newcolumntype{M}{>{\centering\arraybackslash}m{1cm}}

\makeatletter
\def\blfootnote{\xdef\@thefnmark{}\@footnotetext} 
\makeatother

\defcitealias{SVR2006}{SVR (2006)}
\addtocontents{toc}{\protect\thispagestyle{empty}}

\graphicspath{ {./figures/} }

\title{Is There a Secular Decline in Disruptive Patents? Correcting for Measurement Bias}

\author{Jeffrey T. Macher\thanks{Corresponding author. Robert E. McDonough School of Business, Georgetown University, 37th and O Streets NW, Washington, DC 20057, USA; \href{mailto:jeffrey.macher@georgetown.edu}{jeffrey.macher@georgetown.edu}.},~~Christian Rutzer\thanks{Faculty of Business and Economics, University of Basel, Peter Merian-Weg 6, 4002 Basel, Switzerland; \href{mailto:christian.rutzer@unibas.ch}{christian.rutzer@unibas.ch}.},~~and Rolf Weder\thanks{Faculty of Business and Economics, University of Basel, Peter Merian-Weg 6, 4002 Basel, Switzerland; \href{mailto:rolf.weder@unibas.ch}{rolf.weder@unibas.ch}.}}

\date{\today}

\begin{document}
\maketitle

\begin{abstract} 
\noindent Despite tremendous growth in the volume of new scientific and technological knowledge, the popular press has recently raised concerns that disruptive innovation is slowing. These dire prognoses were driven in part by \cite{park2023papers}, a \textit{Nature} publication that uses decades of data and millions of observations coupled with a novel quantitative metric (the CD index) that characterizes innovation in science and technology as either consolidating or disruptive. We challenge the \cite{park2023papers} patent findings, principally around concerns of truncation bias and exclusion bias. We show that 88 percent of the decrease in the average CD index over 1980-2010 reported by the authors can be explained by their truncation of all backward patent citations before 1976. We also show that this truncation bias varies by technology class. We further account for a change in U.S. patent law that allows for citations to patent applications in addition to patent grants---something ignored by the authors in their analysis---and update the analysis to 2016. We show that the number of highly disruptive patents has increased since 1980---particularly since 2008. Our results suggest caution in using the \cite{park2023papers} patent findings and conclusions as a basis for research and decision-making in public policy, industry restructuring or firm reorganization aimed at altering the current innovation landscape.
\end{abstract}


\vspace*{0.5cm}

\section{Introduction}

Continued innovation in science and technology is considered a bedrock for and driving force of growth and prosperity in most economies. A co-authored paper by Park, Leahey and Funk entitled, “Papers and Patents are Becoming Less Disruptive Over Time,” was recently published in \textit{Nature} (2023). Given its title and findings, as well as the topic examined, the paper attracted significant and global media attention. \cite{park2023papers} has been featured in hundreds of international newspapers and magazines \cite[]{altmetric}. As examples, \cite{economist2023} emphasized in a report on the changing nature of science that "Papers and Patents are Becoming Less Disruptive"; \cite{nytimes2023} asked "What Happened to All of Science's Big Breakthroughs?"; and the \cite{fy2023} noted "Science is Losing its Ability to Disrupt".
 
 One reason for this attention is that the \cite{park2023papers} results show marked declines in disruptive research and innovation over time---a finding complementary to research productivity concerns expressed by other authors \cite[][]{chu2021, Bloom2020, boeing2020}.\footnote{\cite{Bloom2020} conclude in their own analysis that "[t]aking the US aggregate number as representative, research productivity falls in half every 13 years: ideas are getting harder and harder to find" (p. 1138).} \cite{park2023papers} emphasize in their abstract that the "results suggest that slowing rates of disruption may reflect a fundamental shift in the nature of science and technology" (p. 138), and note in their conclusion that "this trend is unlikely to be driven by changes in citation practices or the quality of published work. Rather, the decline represents a substantive shift in science and technology, one that reinforces concerns about slowing innovative activity. We attribute this trend in part to scientists' and inventors' reliance on a narrower set of existing knowledge" (p. 142). 

 A second reason for this attention is that the \cite{park2023papers} results suggest implications for the organization of the entire science and technology innovation process---from government research labs and universities to private and public enterprises. \cite{park2023papers} even provide suggestions as to how universities and federal agencies can "strongly reward research quality" and help scholars to "inoculate themselves from the publish or perish culture, and produce truly consequential work" (pp. 143-144), and conclude that "[u]nderstanding the decline in disruptive science and technology more fully permits a much-needed rethinking of strategies for organizing the production of science and technology in the future" (p. 144).
 
 The authors' findings unsurprisingly piqued the interests of researchers, commentators, and journalists on a global scale. Some offered potential explanations, such as increased pressures to apply for large (interdisciplinary) projects, rising administrative burdens, declining basic research funding, increasing risk-aversion, and heightened publishing pressures \cite[]{rust2023rebooting, yanai2023make}. Other researchers raised concerns around the CD index used by \cite{park2023papers} as an appropriate measure of disruptive innovation \cite[]{petersen2023disruption,leibel2023, Ruan2021,bornmann2020disruption}. An important question arises, however, as to whether the \cite{park2023papers} findings are accurate.
 
 This paper does not challenge the CD index as a measure of disruptiveness of newly created knowledge. The CD index was developed by \cite{funk2017dynamic} and can broadly be understood as a measure of whether a patent (scientific paper) is more \textit{Consolidating} or \textit{Disruptive}, based on the characteristics of its citations. A patent (scientific paper) is disruptive if "the subsequent work that cites it is less likely to also cite its predecessors" \cite[][p. 139]{park2023papers}. Our concerns instead arise from how the authors truncate cited references (i.e., "backward citations") to patents and exclude citations to patent applications. We thus challenge their methodological approach due to truncation bias and exclusion bias. 

Restricting our analysis to the same patent data, we find that the worrisome result suggested by \cite{park2023papers} is mainly a consequence of the omission of citations to older innovations: the authors truncate all backward patent citations before 1976. We also show that substantive differences in the results occur from not considering patent law changes: the authors exclude all patent application citations published after 2000. Both of these biases have large measurement effects on the CD index and direct implications on the authors' findings and conclusions.  

Correcting for truncation bias substantially decreases the apparent secular decline in patent disruptiveness suggested by \cite{park2023papers}: i.e., our analysis indicates a marginal decrease in the CD index over time. Correcting further for exclusion bias challenges the "remarkable stability" (p. 140) in the number of highly-disruptive patents argument suggested by\cite{park2023papers}: i.e., our analysis indicates an \textit{increase} in the number of highly disruptive patents over time---particularly since 2008.\footnote{This finding differs from \cite{kalyani2022} who uses the appearance of new "bigrams" as an indicator of novel innovation ideas, but is consistent with \cite{braguinsky2023} who observe a sharp increase in "novel" United States Patent and Trademark Office (USPTO) patent applications filed since 2005.}

The rest of this paper is organized as follows. Section 2 compares the \cite{park2023papers} methodology (with truncation of backward citations) to our methodology (without truncation of backward citations) using the same patent data as \cite{park2023papers}. Section 3 illustrates why truncation bias is concerning, and Section 4 assesses its significance. Section 5 updates the data and analysis over 2011-2016 and considers a 1999 patent law change that allows for citations to patent grants and to patent applications to assesses the significance of exclusion bias. Section 6 shows that the number of highly disruptive patents has increased over time---particularly since 2008. Section 7 offers concluding remarks and suggests caution against any proposed research policy changes in response to \cite{park2023papers}. 

\section{Measuring Patent Disruptiveness}

The CD index was developed by \cite{funk2017dynamic} and used by \cite{park2023papers} to characterize whether a patent or scientific publication is considered more consolidating (i.e., building upon previous research and reinforcing the status quo) or more disruptive (i.e., obsolescing previous research and pushing into new directions).\footnote{Since its introduction by \cite{funk2017dynamic} the CD index has been used in a wide variety of analyses to capture the disruptive content of patents \cite[e.g.,][]{frey_2023, kaltenberg2023, wu2019} and of scientific papers \cite[e.g.,][]{frey_2023, bornmann2020disruption, wu2019}.} The CD index ranges from -1 (consolidating) to 1 (disruptive). \cite{park2023papers} use five-year post-publication windows in its construction, referred to as $CD_5$, and their analysis starts in 1945 for scientific papers and in 1980 for patents.

\begin{figure}[!ht]
\centering
  \caption{The CD index: \cite{park2023papers} Methodology Versus Our Methodology}

  \includegraphics[width=0.6\linewidth,keepaspectratio]{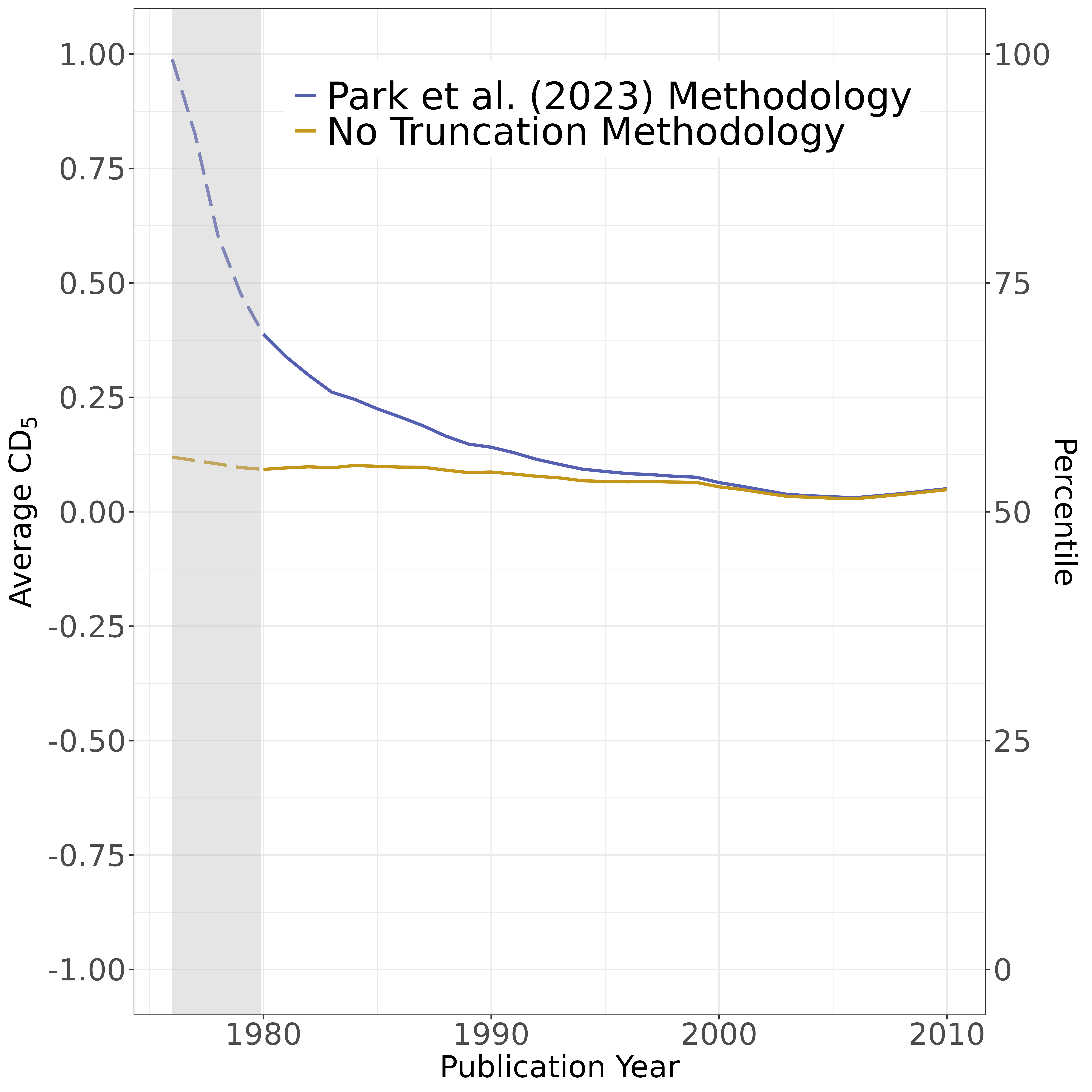}
\label{fig:plot_average_cd}

\begin{minipage}[b]{\textwidth}
\vspace*{0.5cm}
      \footnotesize
    The figure shows the average $CD_5$ Index calculated using the \cite{park2023papers} methodology and our methodology. The \cite{park2023papers} methodology (in blue) omits backward citations to granted USPTO patents published before 1976; our methodology (in yellow) includes all backward citations to granted USPTO patents.  
    \end{minipage}
\end{figure}

Fig. \ref{fig:plot_average_cd} plots two average $CD_5$ indices for approximately $3.66$ million USPTO patents over $1980$-$2010$: one index replicates the \cite{park2023papers} methodology; the other index uses our methodology.\footnote{As in \cite{park2023papers}, we consider only utility patents and draw from PatentsView (version February $21$, $2023$). This database contains all US patents published between $1976$ and September $29$, $2022$. While \cite{park2023papers} focus on patents of the aggregate NBER technology fields "Chemical", "Computer and Communications", "Pharmaceutical and Medical", "Electrical and Electronic", and "Mechanical" that encompasses $3,046,672$ granted utility patents over 1980-2010, our main analysis utilizes all granted USPTO utility patents published that encompasses $3,662,051$ granted utility patents over the same period. See Appendix \ref{data} for more details.} As is readily apparent, the evolution of each index differs markedly: the \cite{park2023papers} methodology produces an average annual CD index that starts at 0.39 in 1980 (the first observation year) but declines rapidly to 0.05 in 2010; our methodology instead produces an average annual CD index that starts at 0.09 in 1980 and declines more gradually to 0.05 in 2010. Fig. \ref{fig:plot_average_cd} nonetheless shows marked convergence in the two indices in the later sample years---particularly since 2000. 

We express changes in these respective CD indices by using percentiles across their full range: i.e., -1 (0th percentile), 0 (50th percentile), and 1 (100th percentile) as shown on the right vertical axis of Fig. \ref{fig:plot_average_cd}. The \cite{park2023papers} methodology indicates the CD index declines from the 69.5 percentile in 1980 to the 52.5 percentile in 2010: a substantive decrease of 17 percentage points relative to its maximum possible change.\footnote{The percentile of a CD index value can be calculated by 100*(CD index/2 + 0.5).} Our methodology indicates the CD index declines from the 54.5 percentile in 1980 to the 52.5 percentile in 2010: a marginal decrease of two percentage points relative to its maximum possible change. 

The contrast in these CD indices would be even greater had \cite{park2023papers} fully exploited the patent data available to them over 1976-2017. The authors indicate that they end their analysis before 2017 "because some measures require data from subsequent years for calculation" (p. 145): i.e., a five-year post-publication window.\footnote{This five-year post-publication window does allow, in principle, for an average $CD_5$ index up to 2012; the authors nevertheless end their analysis in 2010.} For illustration, suppose all of the available patent data were used up to 1976 instead of starting at 1980. The dotted line in the gray-shaded area in Fig. \ref{fig:plot_average_cd} using the \cite{park2023papers} methodology indicates a CD index that declines from the 99.5 percentile in 1976 to the 52.5 percentile in 2010 (i.e., 47 percentage points), whereas our methodology indicates a CD index that declines from the 56.0 percentile to the 52.5 percentile over the same period (i.e., 3.5 percentage points). 

The differences observed in the average CD indices over time are due to \cite{park2023papers}'s truncation of all backward citations to patents published before 1976. Consequently, the results presented by \cite{park2023papers} are biased by this truncation, with the bias strongest for those patents published closer to the truncation year. Their findings of a sharp decline in the average patent disruptiveness can mostly be attributed to a reduction in measurement bias; not to a decline in disruptive innovations. We find that 88 percent of the total decline in the average disruptiveness of patents over 1980-2010 found by \cite{park2023papers} can be explained by the authors' artificial truncation of all backward citations to patents published before 1976.\footnote{The percentage decrease due to truncation is calculated as follows: (1) the $CD_5$ values using the \cite{park2023papers} Methodology are 0.39 in 1980 and 0.05 in 2010 -- a 0.34 difference; (2) the $CD_5$ values using our No Truncation Methodology are 0.09 in 1980 and 0.05 in 2010 -- a 0.04 difference. The difference in differences between the methods is 0.34 - 0.04 or 0.3. Expressed as a percentage: 0.3 / 0.34 * 100 = 88\% of the total decrease in average disruptiveness.}

\section{Why Truncation Matters}

An example helps to illustrate our argument: We compare two patents that both appear in the \cite{park2023papers} dataset:\footnote{Appendix Fig. \ref{fig:pat_example} shows the first page from each patent document listing the backward citations.} US 4181011 was issued in 1980; US 6511791 was issued in 2003. US 4181011 makes 13 citations to patents published between 1958 and 1974; US 6511791 makes 11 citations to patents published between 1987 and 2001. Using the \cite{park2023papers} methodology and replication data  (\textit{$patentsview\_analytical\_df.csv$}), the number of backward citations for patent US 4181011 is zero---given the exclusion of backward citations to patents published before 1976. In contrast, the number of backward citations for patent US 6511791 is 11---the same number as listed in the patent document.

\begin{figure}[!ht]
  \caption{How Truncation of Backward Citations Affects the CD Index}

\centering
\includegraphics[width=0.8\linewidth,keepaspectratio]{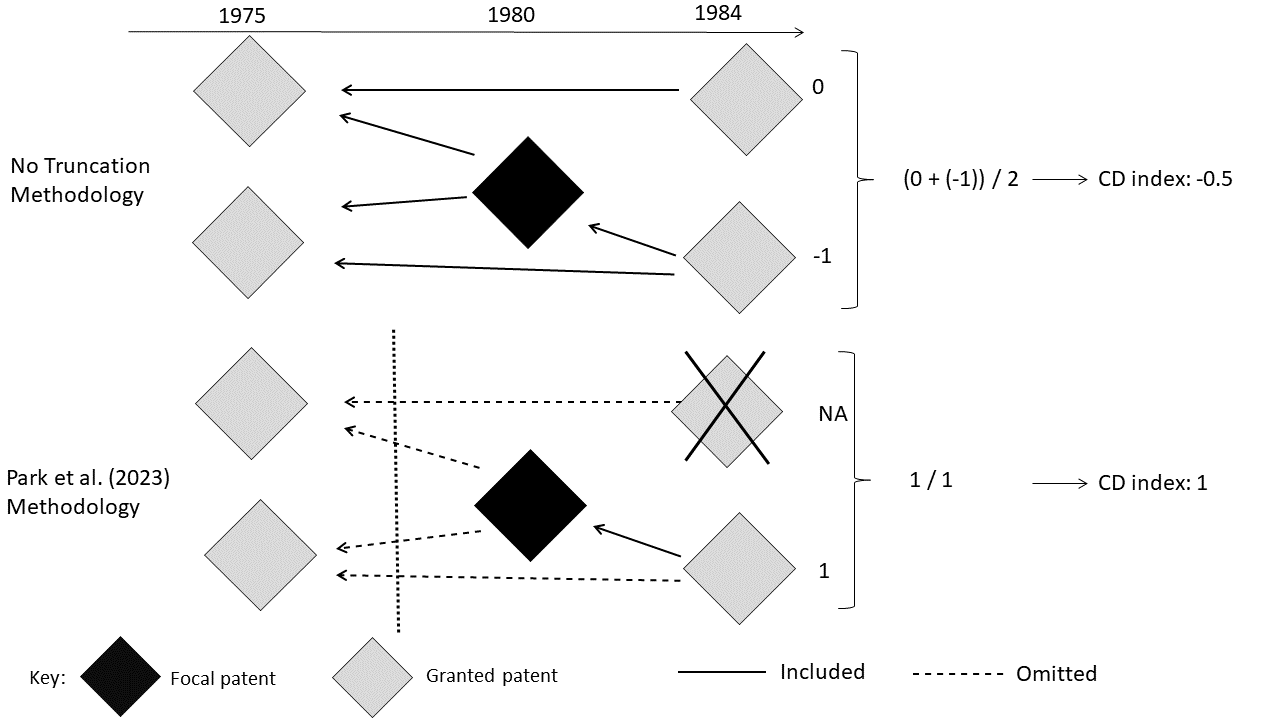}
\label{fig:illustration_truncation}

\begin{minipage}[b]{\textwidth}
\vspace*{0.5cm}

      \footnotesize
    The illustration shows how the CD index of a patent is biased upwards if backward citations are truncated. 
    \end{minipage}
\end{figure}

Fig. \ref{fig:illustration_truncation} provides intuition as to how the truncation of backward citations at a given point in time can bias the CD index: without truncation, the focal patent has a CD index value of -0.5 and is considered consolidating; with truncation (i.e., excluding pre-1976 backward citations), the focal patent  has a CD index value of 1 and is considered disruptive. 

This can be shown formally as follows. \cite{park2023papers} define the CD index for a focal patent as: 

\begin{equation}\label{cd_index}
    CD_t = \frac{1}{N}\sum_{i = 1}^{N}(-2f_{it}b_{it}+f_{it}),~with
\end{equation}
    
$f_{it}= \begin{cases} 
1, & \text{if $i$ cites the focal patent within $t$ years of post-publication of the focal patent}\\
0, & \text{otherwise}
\end{cases}$

and 

$b_{it}= \begin{cases} 
1, & \parbox[t]{0.8\linewidth}{if $i$ cites parts of the predecessors of the focal patent within $t$ years of post-publication of the focal patent}\\
0, & \text{otherwise}
\end{cases}$

\vspace*{0.5cm}

\noindent where $N$ is the sum of the number of patents citing: the focal patent only ($N_F$), the focal patent and parts of its predecessors ($N_B)$, and parts of its predecessors only ($N_R$) within $t$ years after the focal patent publication. A forward citation increases the consolidating (disruptive) nature of the focal patent if the forward cited patent does (does not) cite the predecessor patents (i.e., those patents also cited by the focal patent).\footnote{\cite{wu2019} and subsequent work by \cite{bornmann2020disruption} simplified the CD index formula as: $CD = \frac{N_F-N_B}{N_F+N_B+N_R}$.}  

We extend the CD index by directly considering truncation, which affects the CD index in two ways. First, left truncation reduces the number of counted forward citations $N$ if the truncated backward citations are cited within the $t$ post-publication years of the focal patent: 

\begin{align}\label{n}
N^{tr}<N^{non-tr}
\end{align}
where $tr$ represents truncated and $non-tr$ represents non-truncated backward citations.

Second, left truncation reduces the $b_{it}$ value if a forward citation of the truncated predecessor is also a forward citation of the focal patent within $t$ years of publication. In this case: 
\begin{align}\label{b}
-f_{it}b_{it}^{tr}=0\text{ and } -f_{it}b_{it}^{non-tr}=-1 \rightarrow -f_{it}b_{it}^{tr}>-f_{it}b_{it}^{non-tr}.
\end{align}

Using equations \eqref{n} and \eqref{b}, we consider four distinct cases: (i) no backward citations are truncated; (ii) backward citations are truncated but none are cited within the $t$ post-publication years of the focal patent; (iii) backward citations are truncated and cited within the $t$ post-publication years of the focal patent but none cite the focal patent; and (iv) backward citations are truncated and cited within the $t$ post-publication years of the focal patent and at least one of them cite the focal patent. In case (i) and case (ii), equations \eqref{n} and \eqref{b} hold as equalities and truncation has no effect on the CD index:

\begin{equation}\label{trunc_equal}
CD_t^{tr} = CD_t^{non-tr}.
\end{equation}

In case (iii), the inequality of equation \eqref{n} holds but equation \eqref{b} is strictly an equality. In case (iv), the inequalities of equations \eqref{n} and \eqref{b} hold. In case (iii) and case (iv), the truncated CD index is biased upward as long as the summation part of equation \eqref{cd_index} of the truncated CD index is positive, because each of the two effects artificially increases its value. Moreover, the bias is stronger in case (iv), \textit{ceteris partibus}. Both results are easily seen by adding the inequalities of equations \eqref{n} and \eqref{b} to the CD index provided in equation \eqref{cd_index}. As a result:
\begin{equation}\label{trunc_larger}
CD_t^{tr} > CD_t^{non-tr}.
\end{equation}

Truncation may therefore lead to upward or downward bias in the CD index if the summation of equation \eqref{cd_index} is negative and the inequalities of either equation \eqref{n} or equations \eqref{n} and \eqref{b} hold. In this case:
\begin{equation}\label{trunc_larger_smaller}
CD_t^{tr} \gtrless  CD_t^{non-tr}.
\end{equation}

\section{Assessing Truncation Effects}

\begin{figure}[!ht]
\caption{Truncation Time is Not Important}

\centering
\includegraphics[width=0.45\linewidth,keepaspectratio]{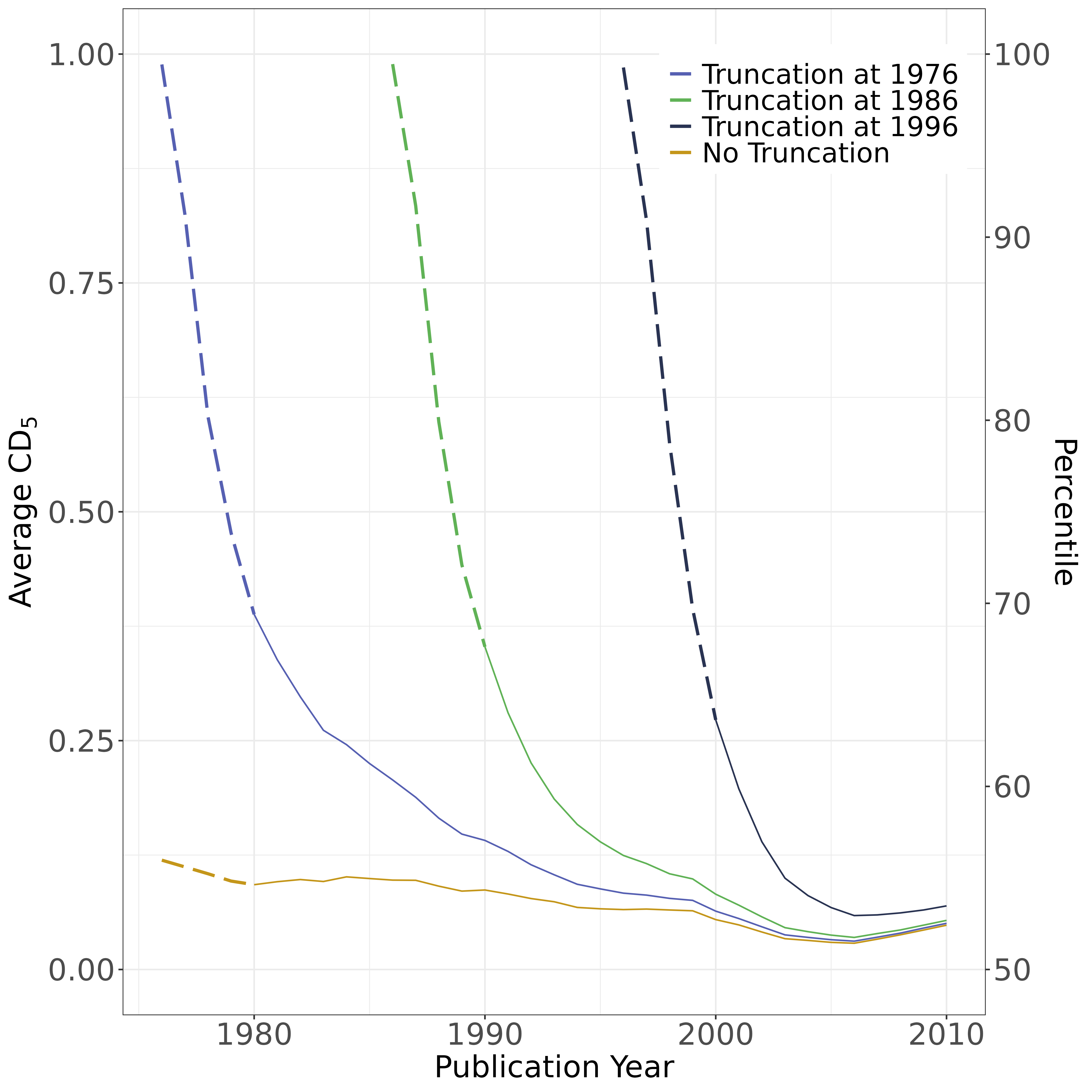}
\label{fig:plot_diff_trunc}

\begin{minipage}[b]{\textwidth}
\vspace*{0.5cm}
      \footnotesize
    The figure shows average $CD_5$ indices excluding backward citations to patents published before 1976 (blue line) as in \cite{park2023papers}, before 1986 (green line), and before 1996 (black line). \cite{park2023papers} use a four-year window from the truncation year for the start of their analysis. The dashed lines show how the truncated $CD_5$ indices change post-truncation year and within this four-year window for illustrative purposes. The solid lines show how the $CD_5$ indices change post-truncation year and beyond this four-year window \cite[as examined by][]{park2023papers}. The average $CD_5$ index with no truncation is included for comparison purposes.
    \end{minipage}
\end{figure}

It is readily apparent that the number of years considered for backward citations matters. The closer to (further from) the truncation year, the greater (lesser) the share of patents with no backward citations and, thus, the larger (smaller) the CD index. As shown in Fig. \ref{fig:plot_average_cd}, had the \cite{park2023papers} methodology started in 1976 (i.e., the truncation year), the bias would be larger: viz., nearly all patents from 1976 would be classified as disruptive. Fig. \ref{fig:plot_diff_trunc} verifies this relationship using different truncation times and demonstrates that the starting year decision for the CD index does not alleviate truncation bias concerns: had the analysis started, e.g., in 1990 (green line) or in 2000 (black line), a four-year truncation window for backward citations would produce similarly sharp CD index declines in the earliest post-truncation years which then stabilize and converge to the unbiased index over time.
 
\begin{figure}[!ht]
  \caption{Size of Truncation Bias Differs by Technologies}
\centering

  \includegraphics[width=1\linewidth,keepaspectratio]{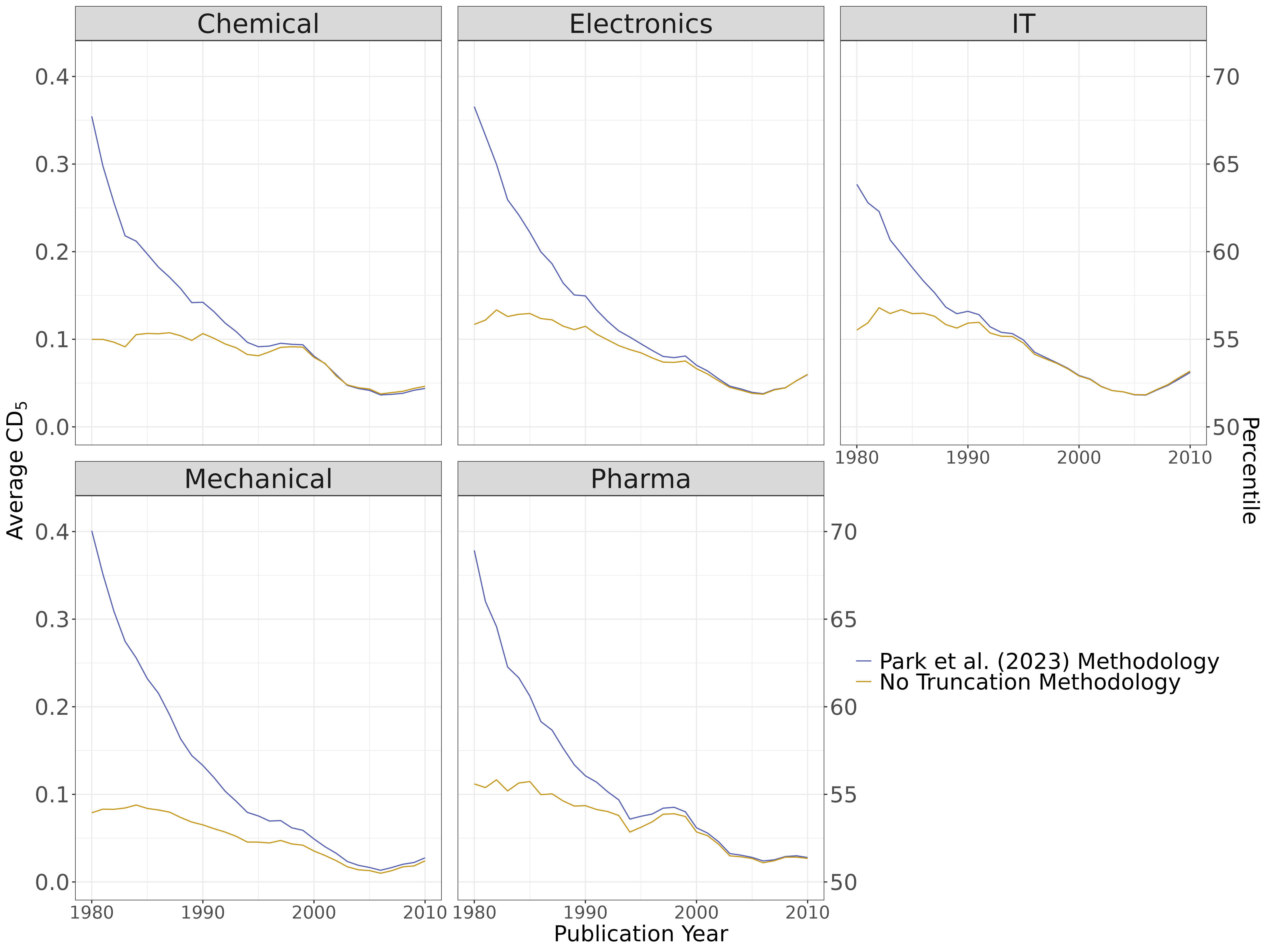}

\label{fig:plot_diff_tech}

\begin{minipage}[b]{\textwidth}
\vspace*{0.5cm}

      \footnotesize
    The figure shows average $CD_5$ indices for various aggregate WIPO technologies, as determined in Appendix Table \ref{tab:tech_field}. The blue line shows the CD index when backward citations to granted U.S. patents published before 1976 are excluded, per the \cite{park2023papers} methodology. The yellow line shows the $CD_5$ index when all backward citations are included, per our methodology.
    \end{minipage}
\end{figure}

Fig. \ref{fig:plot_diff_tech} indicates that this bias is not identical across the technologies examined by \cite{park2023papers}: viz., it is most pronounced in "Mechanical" and least pronounced in "IT". The difference is again most likely explained by truncation: IT patents have developed more recently and rapidly in comparison to Mechanical patents, suggesting older patents are relatively more often cited in the latter category than in the former category (see Appendix Fig. \ref{fig:avg_age_bwc}). As a result, less (more) backward citations are truncated in IT (Mechanical) patents. For comparison, the yellow line represents the $CD_5$ index when all backward citations are included (no truncation) and implies relatively small changes over time and relatively minor differences between technologies.

\section{Updating Years and Patent Law Change}

We examine two factors that could affect the post-2010 CD index. First, we update the patent data analysis window to 2021 to extend the $CD_5$ index to 2016. Second, we consider a major U.S. patent law change: in particular, the Inventor Protection Act of 1999 requires that U.S. patent applications be published 18 months after the initial application filing---effectively shifting citation patterns from strictly patent grants to patent grants and patent applications \cite[]{johnson2003forced}.\footnote{See Appendix \ref{data} for more detailed discussion on the additional data used, and Appendix \ref{change_patent_law} for more detailed discussion of how excluding citations to patent applications affects the CD index.} 

\begin{figure}[!ht]
  \caption{Patent Law Change Creates Exclusion Bias in \cite{park2023papers} Methodology}

\centering

  \includegraphics[width=0.6\linewidth,keepaspectratio]{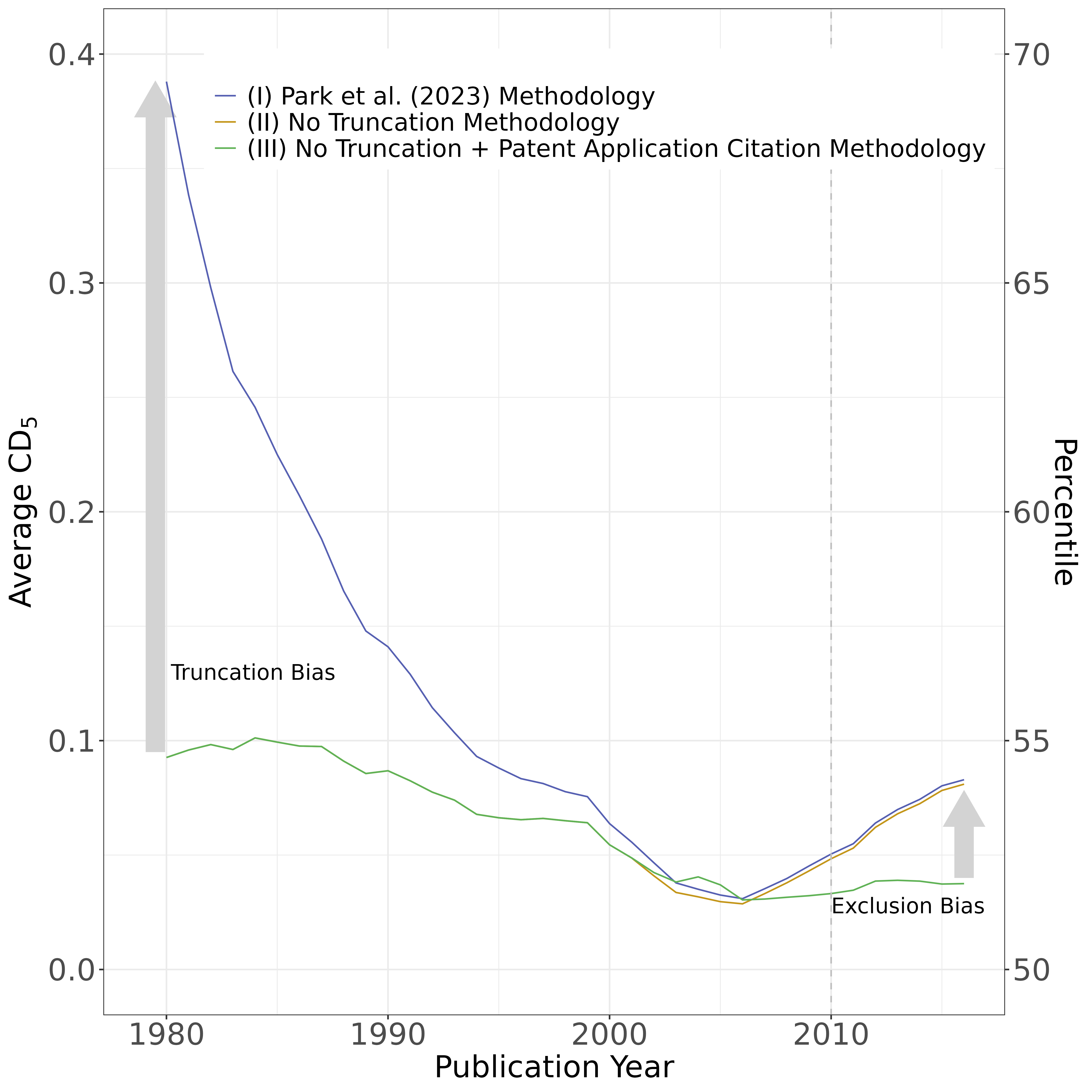}
\label{fig:plot_average_cd_correct}

\begin{minipage}[b]{\textwidth}
      \footnotesize
      \vspace*{0.5cm}

    The figure shows the average $CD_5$ index under different methodologies: (I) the \cite{park2023papers} methodology that excludes backward citations to patent grants before 1976; (II) our methodology that includes all backward citations to strictly patent grants; and (III) our methodology that includes (II) and all backward citations to patent grants and patent applications. The 2010 vertical line denotes the end of the \cite{park2023papers} analysis.
    \end{minipage}
\end{figure}

Fig. \ref{fig:plot_average_cd_correct} shows these post-2010 changes in three average $CD_5$ indices using the same data: (I) the \cite{park2023papers} methodology; (II) our methodology including all backward citations to patent grants only; and (III) our methodology including all backward citations to patent grants and patent applications. Note that methodology (III) considers only those citations to patent applications that receive patent grants within five years after the publication year of a focal patent, as this approach best achieves consistency between the pre- and post-Inventor Protection Act periods.\footnote{Appendix Fig. \ref{fig:plot_app_grant_app_all} shows similar results when citations to patent applications that have not (yet) received patent grants are included.}

Fig. \ref{fig:plot_average_cd_correct} is identical to Fig. \ref{fig:plot_average_cd} up to 2001 because---prior to the Inventor Protection Act's passage---patents only cited patent grants. The inclusion of citations before 1976 eliminates the truncation bias and subsequently decreases the CD index over 1980-2005---as seen by comparing methodology (I) to methodologies (II) and (III). The inclusion of citations to patent applications eliminates the exclusion bias and subsequently decreases the CD index over 2005-2016---as seen by comparing methodologies (I) and (II) to methodology (III). Correcting for both biases, methodology (III) indicates the average CD index declined slightly from the 54.5 percentile in 1980 to the 52 percentile in 2005, and has been relatively stable since.\footnote{Appendix Fig. \ref{fig:cor_plot} shows the correlation of $CD_5$ values of focal patents, backward citations of focal patents, citations to focal patents, and citations to focal patents and/or their predecessors for the three methodologies over time. The correlation of the $CD_5$ values between the different methods is below 0.5 at the beginning (1980) and end (2016) of our analysis: In 1980, this is mainly due to low correlations for the backward citations of the focal patents; In 2016, this is mainly due to low correlations for citations to the focal patents.} 

Our analyses clearly document that the \cite{park2023papers} CD index patent results are significantly influenced by the time window chosen and the biases present. Extending the patent analysis to 2016, moreover, reveals increasing average CD values---especially since 2005---due to the exclusion of citations to patent applications. In addition, methodology (II) in Fig. \ref{fig:plot_average_cd_correct} shows the CD index evolution if backward citations to pre-1976 patents are considered but backward citations to patent applications are not: a pronounced U-shape results with nearly equal CD index averages in 1980 (start point of the green line) and in 2016 (end point of the yellow line). 

\section{Increasing Number of Highly Disruptive Patents}

\begin{figure}[!ht]
  \caption{Investigation of Highly Disruptive Patents}

\let\nobreakspace\relax
\centering

  \includegraphics[width=1\linewidth,keepaspectratio]{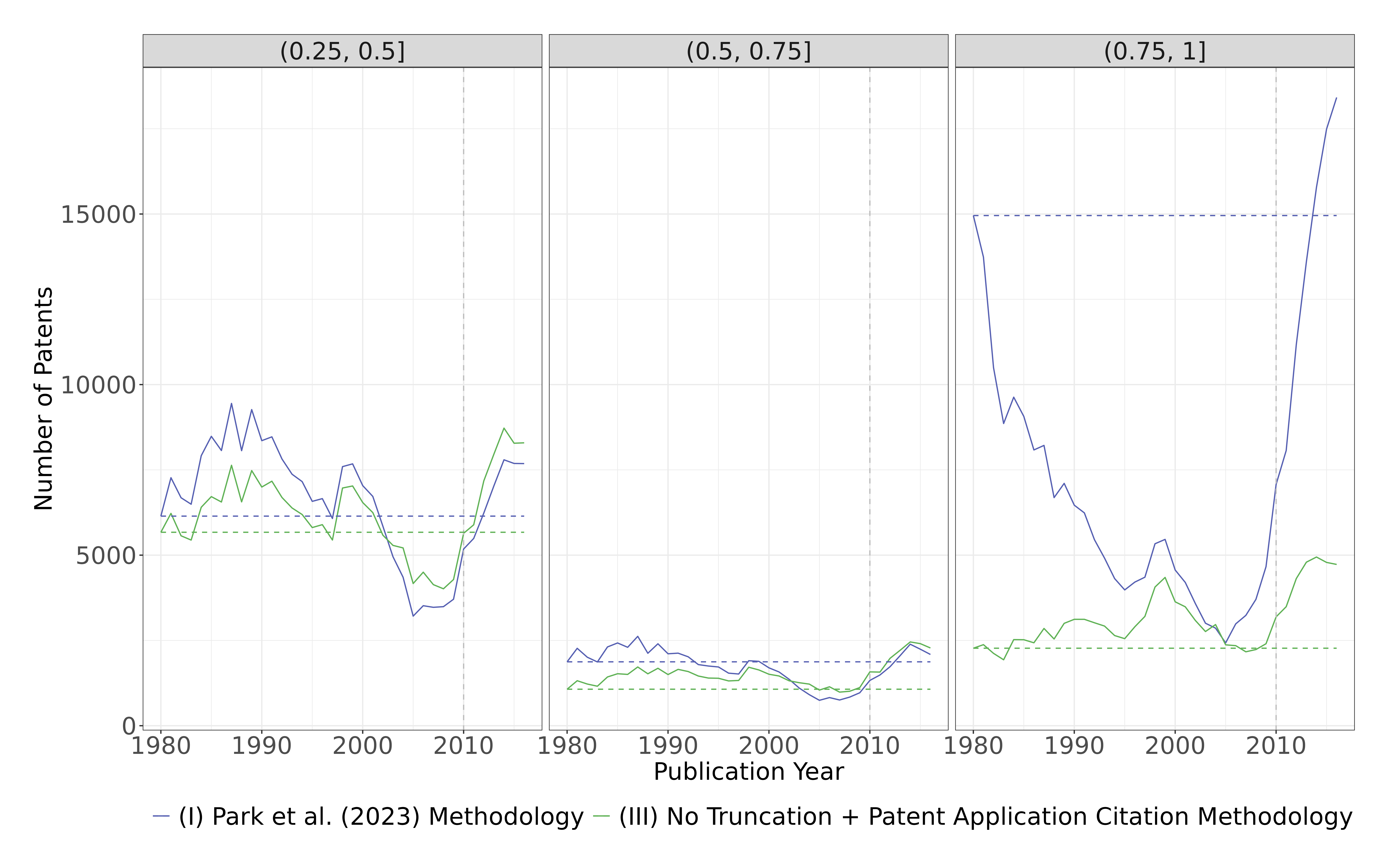}

\label{fig:plot_num_dis_group_diff_high_disr}

\begin{minipage}[b]{\textwidth}
      \footnotesize
      \vspace*{0.5cm}

   The figure shows the number of highly disruptive patents over time for three highly disruptive patent groups: Panel (a) uses $CD_5\in(0.25,~0.5]$; Panel (b) uses $CD_5\in(0.5,~0.75]$; and Panel (c) uses $CD_5\in(0.75,~1]$. Each panel compares the \cite{park2023papers} methodology to our (III) methodology. The respective horizontal lines show the 1980 values for comparison purposes. The vertical line at 2010 represents the end of the \cite{park2023papers} analysis.  
    \end{minipage}
\end{figure}

\cite{park2023papers} emphasize in their analysis that "despite large increases in scientific productivity, the number of papers and patents with $CD_5$ values in the far right tail of the distribution remains nearly constant over time" (p. 140). We evaluate this conclusion with respect to patents---after correcting for truncation bias and exclusion bias---by calculating the number of patents in the three highly disruptive patents groups defined by \cite{park2023papers}: $CD_5\in(0.25,~0.5]$, $CD_5\in(0.5,~0.75]$, and $CD_5\in(0.75,~1]$. Fig. \ref{fig:plot_num_dis_group_diff_high_disr} illustrates the results for each patent group in separate panels. In contrast to \cite{park2023papers}, our analysis reveals a notable \textit{increase} in the number of highly disruptive patents over time.\footnote{Appendix Figure \ref{fig:plot_diff_tech_grant_appl} shows the number of very highly disruptive patents for five aggregated WIPO technologies. Using the \cite{park2023papers} methodology, the absolute number of very highly disruptive patents over 1980-2010 has notably decreased in Chemical and Mechanical; moderately increased in Pharmaceuticals; and notably increased in Electronics and IT. Our Methodology III instead indicates increases in the absolute number of very highly disruptive patents over 1980-2010 in all WIPO technologies except Mechanical.}

Where do the differences come from? Fig. \ref{fig:plot_num_dis_group_diff_high_disr} illustrates that the \cite{park2023papers} methodology results in declining CD indices over 1980-2010 in each highly disruptive patent group---particularly for $CD_5\in(0.75,~1]$.\footnote{In particular, the absolute number of patents in 2010 was 84\% of the patents in 1980 in $CD_5\in(0.25,~0.5]$, 71\% in $CD_5\in(0.5,~0.75]$, and 47\% in $CD_5\in(0.75,~1]$.} A sensible conclusion based on their methodology is that the number of highly disruptive patents declined over 1980-2010.\footnote{One potential reason why the authors suggest "remarkable stability in the absolute number of highly disruptive works" (p. 140) may be that Figure 4b (p. 141) in \cite{park2023papers} also shows $CD_5\in(0,~0.25]$), which includes a substantially larger number of patents but does not belong in the highly disruptive category (according to their definition). As a result, the vertical axis masks the declines in the other three patent groups.} Once truncation bias and exclusion bias are corrected for, however, Fig. \ref{fig:plot_num_dis_group_diff_high_disr} reveals that the number of highly disruptive patents increased over 1980-2010 in all three patent groups.\footnote{In all three panels, the absolute number of highly disruptive patents differs from that reported in Fig. 4b (p. 141) of \cite{park2023papers} because the authors only consider a subset of technologies (see Appendix \ref{data}). Apart from this difference, the shape of the change is the same.} Extending the data to 2016, moreover, indicates that the number of very highly disruptive patents more than doubled and the number of patents in the other disruptive patent groups increased significantly. This analysis also confirms Methodology (I) distortions---especially for very highly disruptive patents---as shown in the right panel of Fig. \ref{fig:plot_num_dis_group_diff_high_disr}: the left-side of the U-shape arises from truncation bias (i.e., no citations to pre-1976 patent grants); the right-side arises from exclusion bias (i.e., no citations to patent applications).

Appendix Table \ref{tab:conv_matrix_trunc} examines truncation bias in more detail. This table shows how patents in 1980 that belong to a particular $CD_5$ category using the \cite{park2023papers} methodology (with truncation bias) are distributed across $CD_5$ categories using our methodology (without truncation bias). Backward citation truncation of pre-1976 patents misclassifies a relatively high proportion of patents as highly disruptive. For example, 25 percent of the USPTO patents in 1980 that are considered very highly disruptive with truncation bias are actually consolidating without truncation bias. 

Appendix Table \ref{tab:conv_matrix_appl} examines exclusion bias in more detail. This table shows how patents in 2016 that belong to a particular $CD_5$ category using the \cite{park2023papers} methodology (with exclusion bias) are distributed across $CD_5$ categories using our methodology (without exclusion bias). Backward citation exclusion to patent applications similarly misclassifies a relatively high percentageof recent patents as highly disruptive. For example, 19 percent of the USPTO patents in 2016 that are considered very highly disruptive using the \cite{park2023papers} methodology are actually consolidating. Determining the disruptive potential of individual patents thus requires careful consideration to ensure against measurement bias and misclassification.\footnote{Note that our findings also affect variations of the index proposed in \cite{park2023papers} which are partly based upon other research \cite[]{leydesdorff2021proposal,bornmann2020disruption}. In Appendix Fig. \ref{fig:cd_diff_time} and \ref{fig:plot_norm}, we show the time series for several variations of the CD index used by \cite{park2023papers}: e.g., with respect to an extended time-window for forward citations or to a normalization of the index. In all cases, the truncation of backward citations leads to an analogous upward bias as described above.}

\section{Discussion}

Our examination of the \cite{park2023papers} CD index methodology was prompted by the stark decrease in the share of disruptive innovation in science and technology reported by the authors in patents over 1980-2010 and in scientific publications over 1945-2010. Our methodology and analysis of patents suggest results that either question or differ markedly from the \cite{park2023papers} results, in at least four important respects.

First, we find that the secular decline in the average CD index found by \cite{park2023papers} is mainly due to the truncation of backward citations, while a modest increase in the average CD index post-2005 is mainly due to the exclusion of backward citations to patent applications. In comparison to the \cite{park2023papers} methodology, correcting for both biases results in a CD index that starts significantly lower in 1980, modestly declines to 2005, and is largely stable since. A CD index without measurement bias exhibits a marginal decline (viz., two and a half percentage points) but no secular decline (viz., 17 percentage points) in disruptive innovation over 1980-2010, as suggested by the authors. Second, we find that the starting truncation date does not matter. In particular, any backward citation cutoff in any chosen year produces similar CD index patterns as suggested by \cite{park2023papers}. Third, we find that the size of the truncation bias differs markedly across aggregate WIPO technology classes, given differences in the pattern and pace of innovation among these technologies. In particular, it is most pronounced in Mechanical and least pronounced in IT, precisely because more backward citations are truncated in Mechanical and less backward citations are truncated in IT. Fourth, we find that the number of highly disruptive patents has notably increased---particularly since 2008. Once truncation bias and exclusion bias are corrected for, our results challenge \cite{park2023papers} in their assessment of stability in the number of highly disruptive patents over time. 

We suggest that our analysis is important for the research community that may work with biased data or overlook measurement issues---either via the CD index or other related measures. Our analysis highlights, in particular, the dangers of truncation bias from backward citations and exclusion bias from citations to patent applications. Our analysis suggests, perhaps more importantly, caution against overt policy interventions, regulatory changes, or substantive industry and firm reorganizations that seek to improve or disrupt the innovation status-quo based upon the \cite{park2023papers} results. If it was accurate that breakthroughs in patented scientific knowledge have substantially decreased over time, then the current organization of the entire technology innovation process should be reevaluated. In particular, reforms that seek to increase disruptive innovation or restart innovation within R\&D centers and research labs---e.g., in government agencies, universities and firms---would be deemed necessary and paramount. But changes to the current science and innovation process without a solid understanding of the facts in evidence is wrongheaded and might have unintended consequences.

\clearpage 

\section{CRediT authorship contribution statement}
Jeffrey T. Macher: Conceptualization, Writing – original draft, Writing – review \& editing. Christian Rutzer: Calculation, Conceptualization, Investigation, Visualization, Writing – original draft, Writing – review \& editing. Rolf Weder: Conceptualization, Writing – original draft, Writing – review \& editing.

\section{Declaration of competing interest} 
The authors have received no external funding and declare no competing interests.

\section{Acknowledgments}
We thank the editor, Adam B. Jaffe, and three anonymous referees for valuable comments that helped improve this paper.

\section{Data availability}
The code and data needed to reproduce this study are available at Zenodo \url{https://zenodo.org/deposit/8020004}. 
The raw data used in this research was obtained from PatentsView, a platform that provides unrestricted access to detailed information on all granted USPTO patents published since 1976. All raw data can be downloaded at \url{https://patentsview.org/download/data-download-tables}.

We used R version 4.1.2 to perform the computations, mainly using the tidyverse and data.table libraries for data processing and ggplot2 for visualization. The computations were performed using SLURM on the sciCORE cluster at the University of Basel, Switzerland.

\clearpage

\appendix

\section{Data\label{data}}

The study uses data from the February 21, 2023 version of PatentsView, which includes all U.S. patents published between 1976 and September 29, 2022. As in previous research \cite[][]{park2023papers, funk2017dynamic}, we focus only on USPTO patents. We include PatentsView data that assigns each USPTO patent to technology categories as defined by the World Intellectual Property Organization (WIPO) \cite[][]{schmoch2008}. The specific WIPO technologies used to create our aggregated technology groups are provided in Appendix Table \ref{tab:tech_field}. It is important to note that our technology classification differs from that of \cite{park2023papers}, who categorize patents based on NBER technology fields that are no longer available.

In Sections 2-4, we use all granted utility patents over 1980-2010. \cite{park2023papers} examine granted utility patents over this timeframe and in the NBER technology fields of "Chemical," "Computers and Communications," "Drugs and Medical," "Electrical and Electronic," and "Mechanical." Their data includes 3,046,672 patents and 29,777,375 backward citations (see replication data file \textit{$patentsview\_analytical\_df.csv$} of \cite{park2023papers}). We examine granted utility patents over the same timeframe, but cover all technology fields. Our data thus includes 3,662,051 patents and 42,617,016 backward citations in the case of no truncation (i.e., as they appear in the original patent documents) and 35,192,186 backward citations in the case of truncation. 

In Sections 5-6, we expand the dataset over 1980-2016, which results in 5,324,224 granted utility patents and 74,943,850 backward citations. With the passage of the Inventor Protection Act of 1999, patents beginning in November 2000 include backward citations to patent applications and to patent grants. Our methodology (III) includes all backward citations to patent grants and to patent applications that were eventually granted using supplemental PatentsView data. We replace all citations to patent applications with the corresponding patent grant information: i.e., the grant number and publication date. This replacement can result in some backward citations with publication dates later than the citing patent, but the backward citation publication date is not relevant to the $CD_5$ index methodology. What matters are publication dates of forward citations. We do not include patent applications published within five years after the publication year of a focal patent but with corresponding patent grants published later in the analysis. This approach results in 87,880,452 backward citations.

The PatentsView citation table contains patents that cite the same patent using different "publication dates", mainly due to typos in the patent reference lists. We apply the following correction: For each granted USPTO patent, we use the reference list and link each patent mentioned to its original patent document, using the patent number indicated. This provides the publication date for each cited patent as it appears in the original published patent document. We replace the date in the reference list with the date in the original document in all cases where the dates differ. We do the same for citations to patent applications. This correction is only possible for citations to patents published after 1975, as the original pre-1976 patent documents are not available in PatentsView. We therefore cannot correct the date of backward citations to patents published before 1976. We note that backward citation date is irrelevant for the CD index calculation, however, because it is not tied to a specific observation period. An incorrect date could nevertheless be problematic for forward citations where timing is important. Since we compute the CD index only for patents from 1976 onwards (and in most cases from 1980 onwards), only forward citations to patents published after 1975 are relevant. For all these citations, we correct the publication date provided by PatentsView using the procedure described above.

\section{CD Index and Patent Law Change\label{change_patent_law}}
The Inventor Protection Act of 1999 might bias the CD index if citations to patent applications are not considered---especially since patent applications are now increasingly cited \textit{vis-a-vis} patent grants. As \cite{kuhn2020patent} note, citations to patent applications accounted for 25 percent of all citations made by USPTO patents in 2015. 

\begin{figure}[!ht]
  \caption{Excluding Backward Citations to Patent Applications Affects the CD Index}

\let\nobreakspace\relax
\centering

  \includegraphics[width=0.85\linewidth,keepaspectratio]{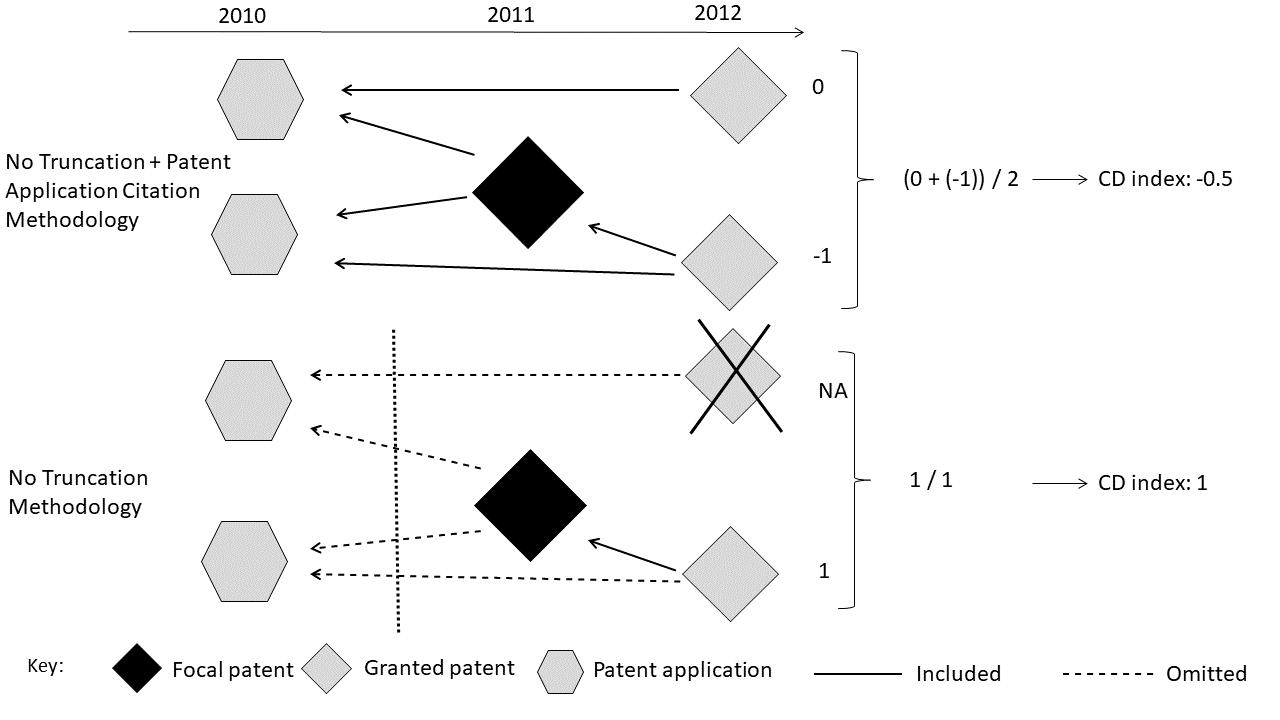}

\label{fig:illustration_omitting_applications}

\begin{minipage}[b]{\textwidth}
      \footnotesize
      \vspace*{0.5cm}

    The illustration shows how excluding citations to patent applications can bias the CD index of a patent, similar to the truncation of backward citations shown in Figure \ref{fig:illustration_truncation}.
    \end{minipage}
\end{figure}

First, as Fig. \ref{fig:illustration_omitting_applications} illustrates, ignoring citations to patent applications can create a bias by falsely declaring a patent as completely disruptive. The channels are the same as when backwards citations are truncated in time. Formally, it affects the CD index via $b_{it}$ and $N$ in equation \eqref{cd_index}, but the driving force is now missing citations to patent applications. As shown by equations \eqref{n}-\eqref{trunc_larger_smaller}, this can create a bias in the CD index similar to left-truncated backward citations.

\begin{figure}[!ht]
  \caption{Backward Citations to Patent Application and Corresponding Patent Grant Affects the CD Index}
\let\nobreakspace\relax
\centering

  \includegraphics[width=0.85\linewidth,keepaspectratio]{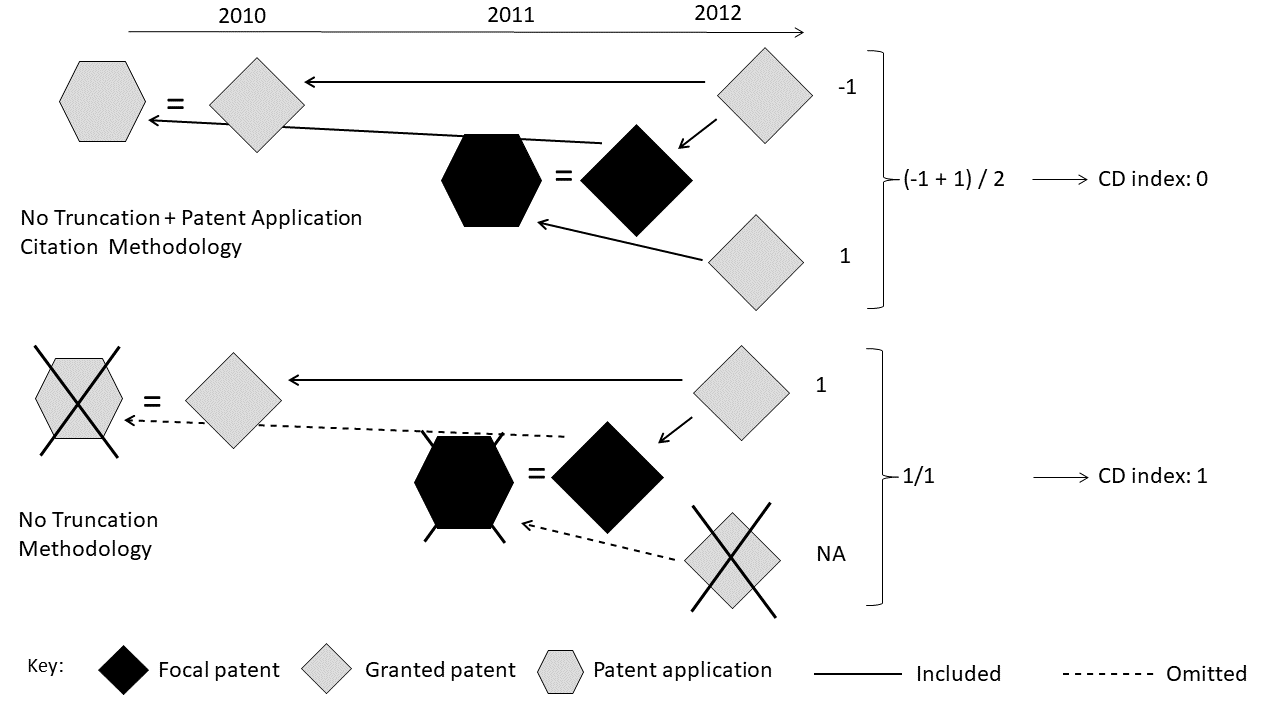}

\label{fig:illustration_omitting_granted_applications}

\begin{minipage}[b]{\textwidth}
\vspace*{0.5cm}

      \footnotesize
    The illustration shows backward citations made to a patent application and the corresponding patent grant can bias the CD index. The patent can be mis-classified as fully disruptive due to the exclusion of citations to the patent application. Specifically, the analysis misses citation to a predecessor application document as well as citation to the application document of the focal patent. 
    \end{minipage}
\end{figure}

Second, ignoring citations to patent applications can create a bias if a patent application is cited in one instance and the subsequent patent grant is cited in another instance. This can occur from citations to predecessor patents or to the focal patent---i.e., applications and grants. Fig. \ref{fig:illustration_omitting_granted_applications} illustrates this possibility in more detail. Formally, it affects the CD index in equation \eqref{cd_index} as before in $b_{it}$ or $N$ but now also in $f_{it}$. This may result in an upward or downward bias of the CD index. An upward bias occurs, for example, when the focal patent cites a patent application and a successor patent cites the focal patent and the patent grant (instead of the patent application). In such a case, the citation to the patent application is missed. A downward bias occurs, for example, if a successor patent only cites the patent application of the focal patent and no predecessor patents (applications or grants) of the focal patent. As this would not be considered a forward citation to the focal patent, it thereby loses some of the disruptive value of the focal patent. The Inventor Protection Act of 1999 may therefore bias the CD index if citations to patent applications for patents published after November 29, 2000, are not properly taken into account.

\clearpage

\subsection{Figures}

\begin{figure}[!ht]
\caption{Examples of Backward Citations in Patent Documents}

\centering
\begin{subfigure}[b]{0.48\textwidth}
         \centering
         \includegraphics[width=\textwidth]{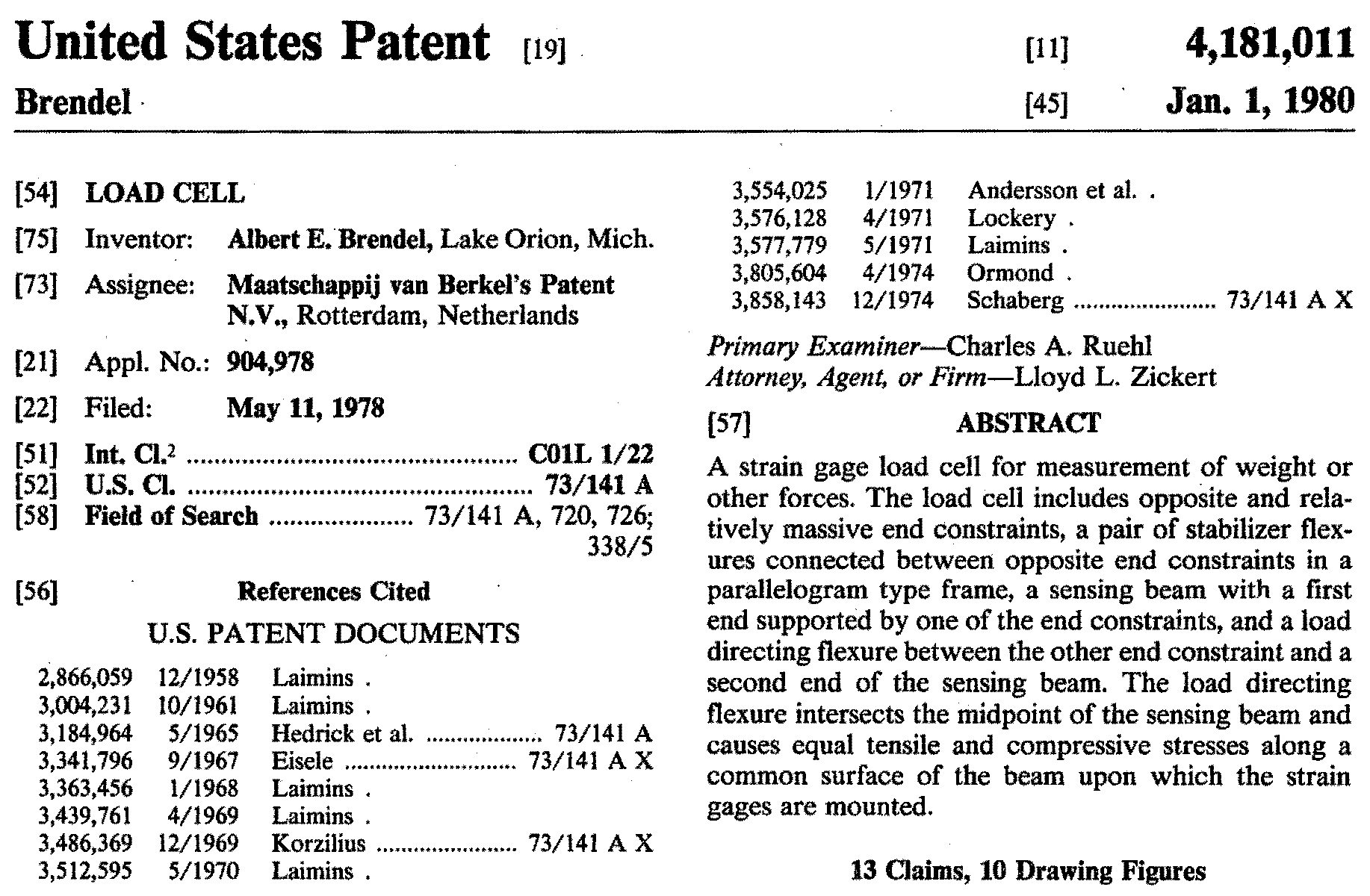}
         \caption{First Page of US Patent 4181011}
         \label{fig:dis_group_right}
     \end{subfigure}
     \begin{subfigure}[b]{0.49\textwidth}
         \centering
         \includegraphics[width=\textwidth]{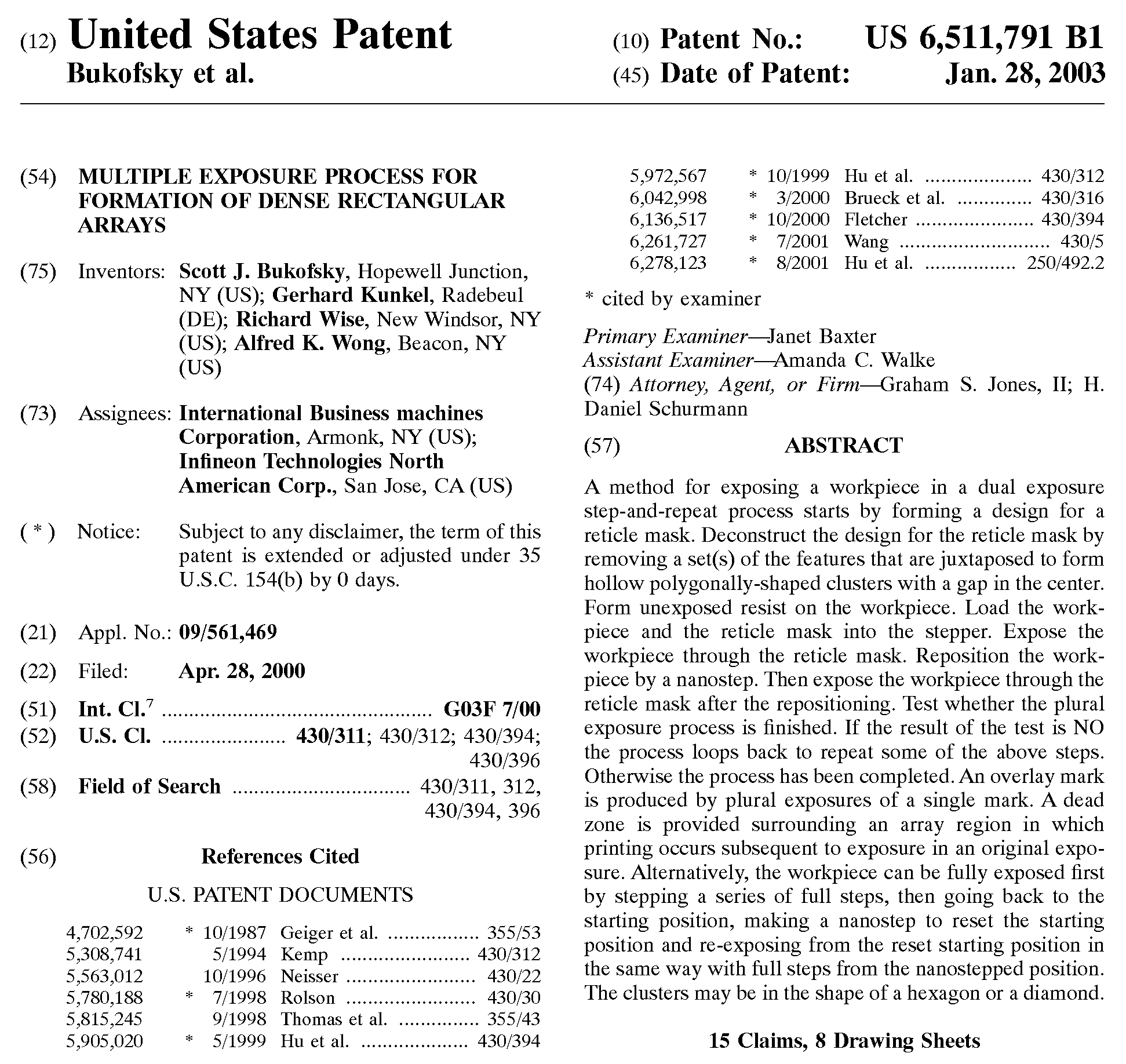}
         \caption{First Page of US Patent 6511791}
         \label{fig:dis_group_left}
     \end{subfigure}
     \hfill

\label{fig:pat_example}

     \begin{minipage}[b]{\textwidth}
\vspace*{0.5cm}

      \footnotesize
    The left panel shows US Patent 4181011 makes 13 backward citations to granted patents, all of which were published before 1976. The right panel shows US Patent 6511791 makes 11 backward citations to other granted patents, all of which were published after 1976.
    \end{minipage}
\end{figure}

\begin{figure}[!ht]
  \caption{Average Backward Citation Age}

\let\nobreakspace\relax
\centering
  \includegraphics[width=1\linewidth,keepaspectratio]{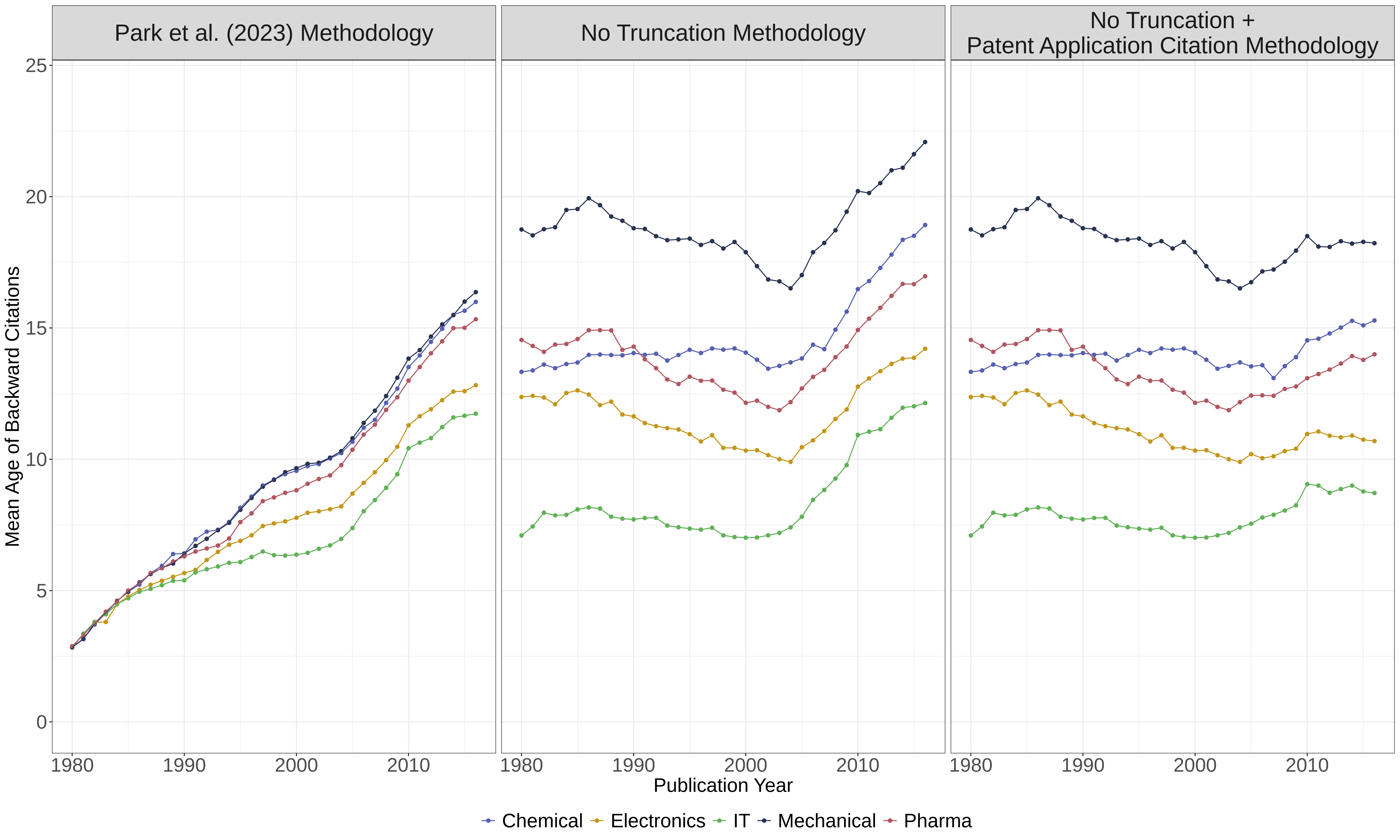}
\label{fig:avg_age_bwc}

\begin{minipage}[b]{\textwidth}
\vspace*{0.5cm}

      \footnotesize
    The figure shows the average age of backward citations for various aggregate WIPO technologies, as determined in Table \ref{tab:tech_field}. The vertical line at 2010 marks the end of the \cite{park2023papers} analysis. The left panel (Truncation Methodology) shows that the average backward citation ages increase in each technology because citations before 1976 are not considered. The middle panel (No Truncation Methodology) shows average backward citation ages are relatively stable in each technology up to 2005 and then increase markedly post-2005 because citations to patent applications are not considered. The right panel (No Truncation + Patent Application Citation Methodology) shows average backward citation ages are relatively stable in each technology and over time.
    \end{minipage}
\end{figure}

\begin{figure}[!ht]
  \caption{Average CD Index By Various Methods}

\let\nobreakspace\relax
\centering

  \includegraphics[width=0.6\linewidth,keepaspectratio]{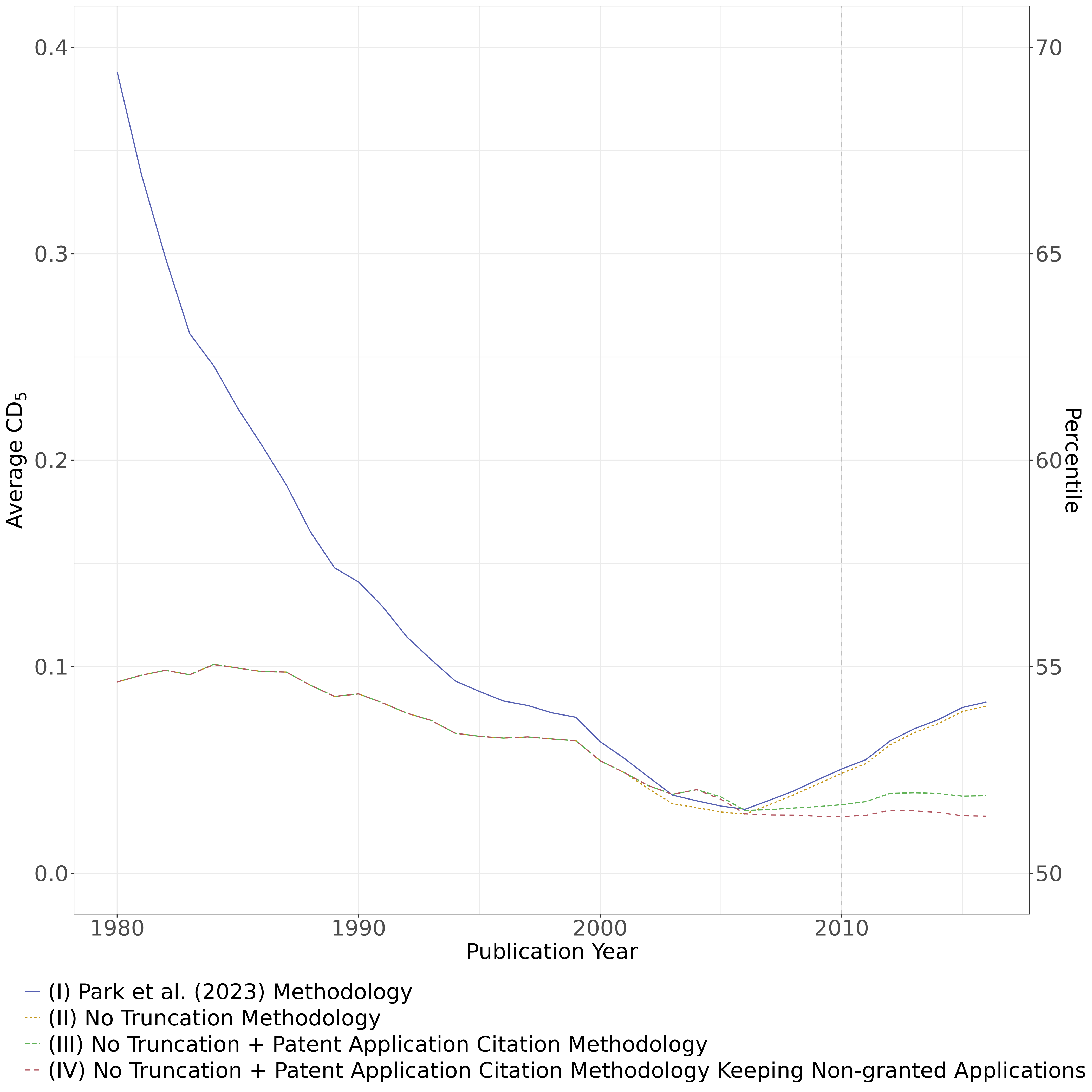}
\label{fig:plot_app_grant_app_all}

\begin{minipage}[b]{\textwidth}
\vspace*{0.5cm}

      \footnotesize
    The figure shows the average CD index calculated using four different methodologies. The vertical line at 2010 marks the end of the \cite{park2023papers} analysis. Methodology (I) via \cite{park2023papers} excludes backward citations to patent grants before 1976 and backward citations to patent applications. Methodology (II) includes backward citations to patent grants before 1976, but excludes backward citations to patent applications. Methodology (III) includes backward citations to patent grants before 1976 and to patent applications that have received patent grants by the end of 2021. This methodology replaces the application number with the corresponding patent grant number and publication date of the application with the publication date of the patent grant. Finally, methodology (IV) includes backward citations to patent grants before 1976, to patent applications that have received patent grants by the end of 2021, and to patent applications that have not received patent grants by the end of 2021. This methodology replaces the application numbers with their corresponding granted patent numbers while keeping the original application dates, which is necessary to ensure that all citations are taken into account as illustrated by Fig. \ref{fig:illustration_omitting_granted_applications}. Note methodology (IV) lies slightly below methodology (III) after 2005 due to the inclusion of backward citations to patent applications that did not receive patent grants and thus would not appear prior to the Act's passage. Citations to patent applications can increase the number of citations to predecessor patents, which in turn may lead to a decrease in the CD index as the value of $N$ in equation \eqref{cd_index} is larger. In this case, however, the decline in the CD index does not necessarily suggest a decline in the number of disruptive innovations. Instead, it could be due to what \cite{petersen2019methods} terms "citation inflation". 
    \end{minipage}
\end{figure}

\begin{figure}[!ht]
  \caption{Correlation Over Time Between Different Variables}

\let\nobreakspace\relax
\centering

  \includegraphics[width=1\linewidth,keepaspectratio]{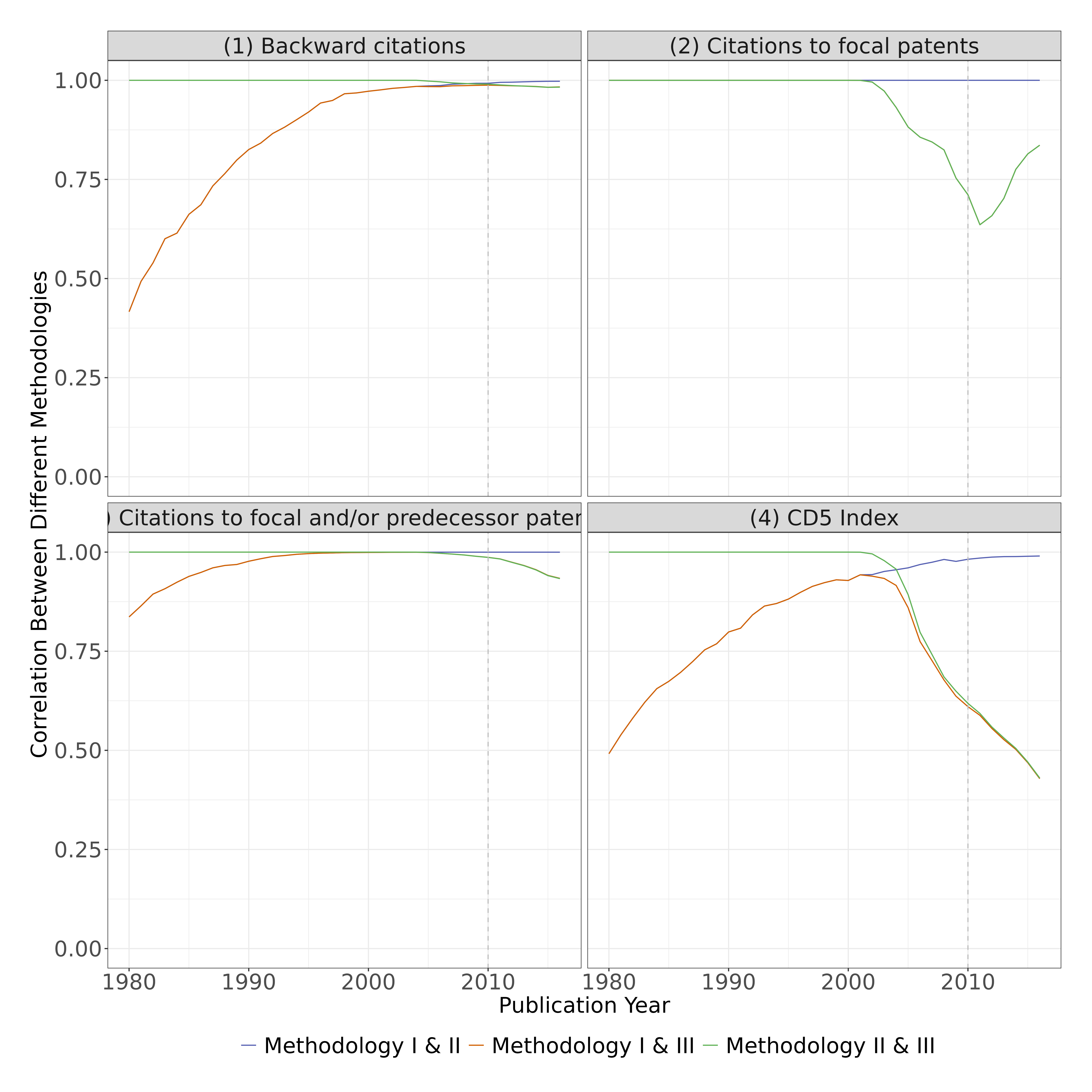}
\label{fig:cor_plot}

\begin{minipage}[b]{\textwidth}
\vspace*{0.cm}
      \footnotesize
    The figure shows correlations between different methods over time for (1) backward citations of focal patents, (2) citations to focal patents (5 years after publication of the focal patent), (3) citations to focal patents and/or their predecessor patents (5 years after publication of the focal patent), (4) and the $CD_5$ index. The blue lines show the correlations between Methodology I and II (Park et al., 2023, Methodology and No Truncation Methodology), the orange line between I and III (Park et al., 2023, Methodology and No Truncation + Patent Application Citation Methodology), and the green line between II and III (No Truncation Methodology and No Truncation + Patent Application Citation Methodology). The vertical line at 2010 marks the end of the \cite{park2023papers} analysis.
    \end{minipage}
\end{figure}

\begin{figure}[!ht]
  \caption{Number of Very Highly Disruptive Patents Across Technologies}
\centering

  \includegraphics[width=1\linewidth,keepaspectratio]{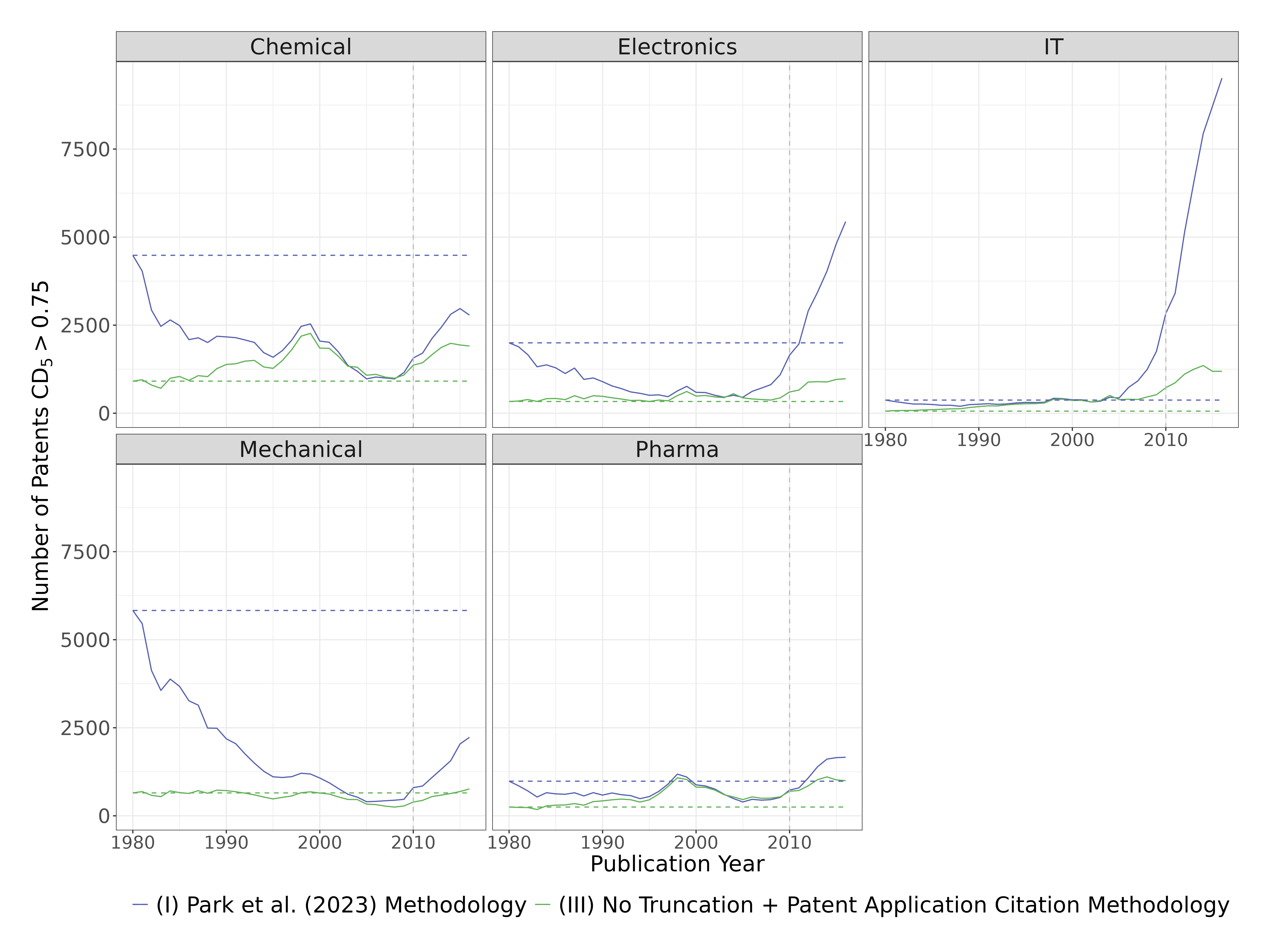}

\label{fig:plot_diff_tech_grant_appl}

\begin{minipage}[b]{\textwidth}
\vspace*{0.5cm}

      \footnotesize
    The figure illustrates the number of very highly disruptive patents for various aggregate WIPO technologies, as determined in Appendix Table \ref{tab:tech_field}. The vertical line at 2010 marks the end of the \cite{park2023papers} analysis. The respective horizontal lines show the values for 1980 for comparison purposes. The blue line shows the number of patents when backward citations to granted U.S. patents published before 1976 are excluded, per the \cite{park2023papers} methodology. The green line shows the number of patents when all backward citations to patent grants and applications are included.
    \end{minipage}
\end{figure}

\clearpage

\begin{figure}[!ht]
  \caption{CD Index With Varying Observation Periods For Forward Citations}

\let\nobreakspace\relax
\centering

  \includegraphics[width=0.9\linewidth,keepaspectratio]{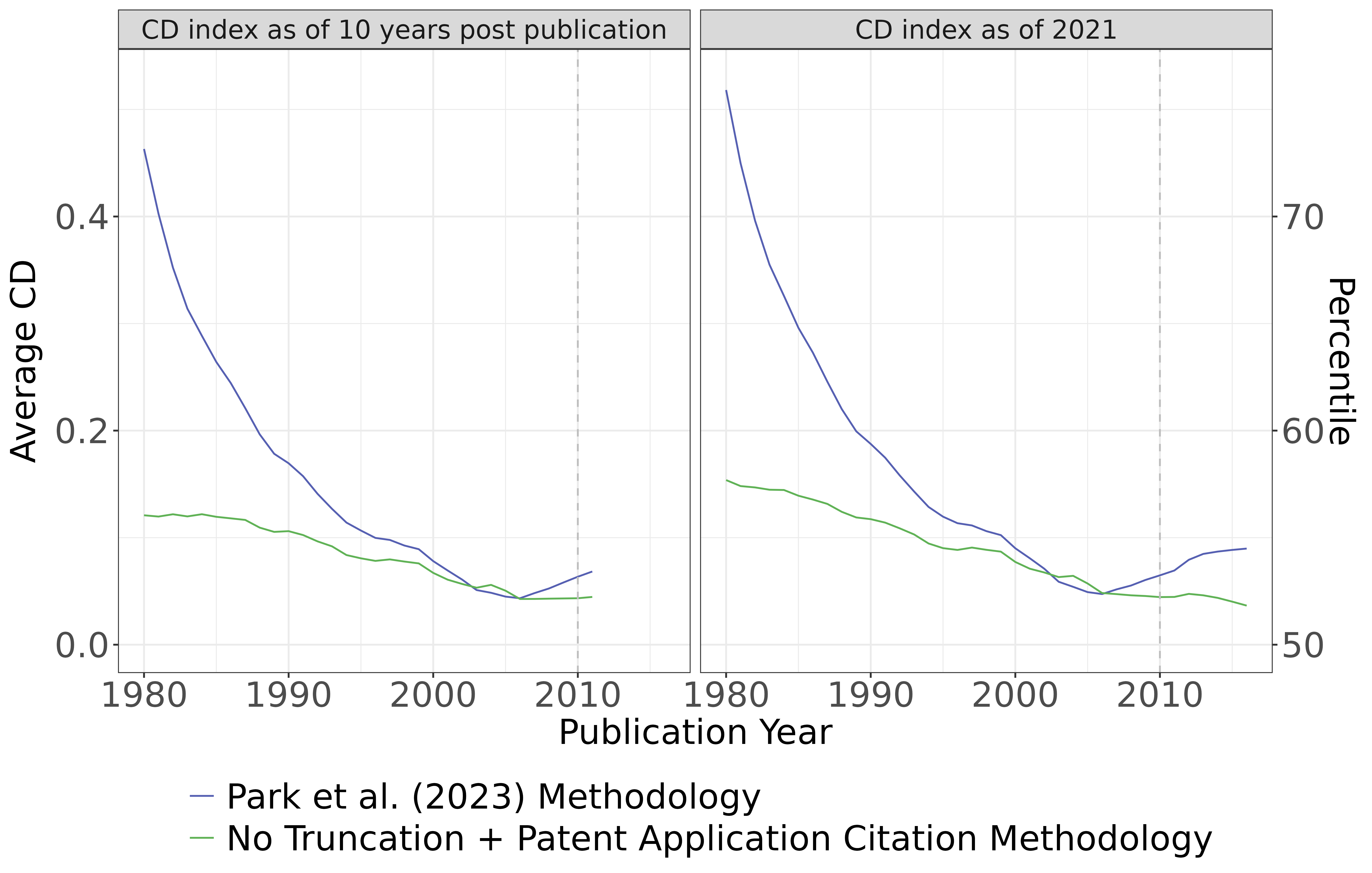}
\label{fig:cd_diff_time}

\begin{minipage}[b]{\textwidth}
\vspace*{0.cm}
      \footnotesize
    The figure shows the average $CD$ Index for different observation periods of forward citations calculated using the \cite{park2023papers} methodology and our methodology that includes all backward citations to patent grants and patent applications. The vertical line at 2010 marks the end of the \cite{park2023papers} analysis. The left-hand panel considers forward citations until 10 years after the publication year of a focal patent. Since data are used until the end of 2021, this value can only be calculated until 2011. The right-hand panel considers all citations until the end of 2021.   
    \end{minipage}
\end{figure}

\begin{figure}[!ht]
  \caption{Different Variants of the CD Index}

\let\nobreakspace\relax
\centering

  \includegraphics[width=1\linewidth,keepaspectratio]{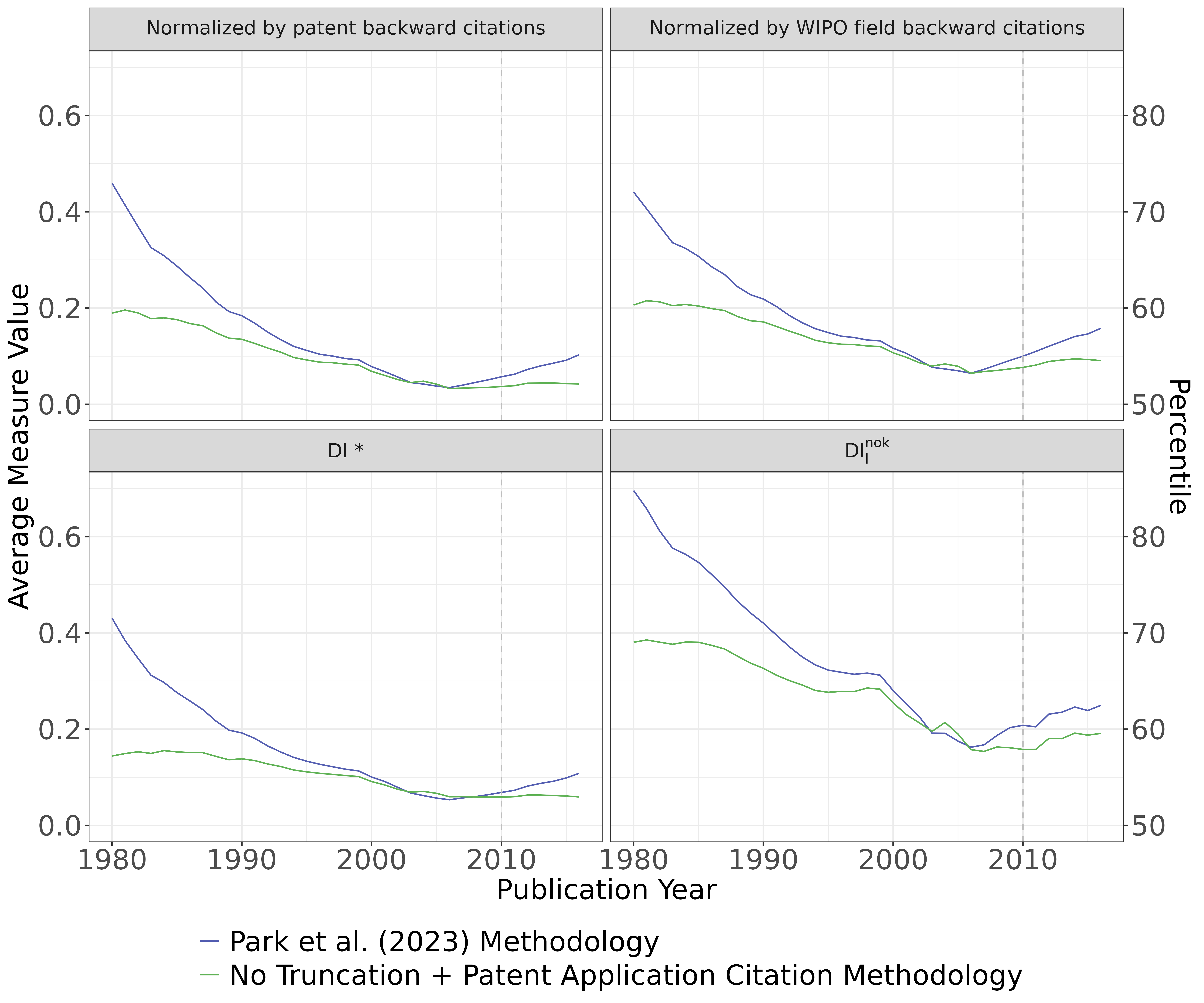}
\label{fig:plot_norm}

\begin{minipage}[b]{\textwidth}
\vspace*{0.cm}
      \footnotesize
 The figure shows different variants of the average $CD$ index as used by \cite{park2023papers}, calculated using the \cite{park2023papers} methodology and our methodology that includes all backward citations to patent grants and patent applications. Each vertical line at 2010 marks the end of the \cite{park2023papers} analysis. The first variant, shown in the upper left panel, normalizes the $CD_5$ index by the number of patents that cite only the predecessors of the focal patents. In particular, the denominator of the CD index (equation \ref{cd_index}) is adjusted by subtracting the number of backward citations of the focal patent: $N_{norm} = N_F + N_B + N_R-N_b$, where $N_F$ is the sum of all citations made only to the focal patent and not to any of its predecessors patents, $N_B$ all citations to the focal patent and parts of its predecessors, and $N_R$ all citations to parts of its predecessors only. In cases where $N_R-N_b<0$, we set $N_R-N_b=0$ as in \cite{park2023papers}. The second variant, shown in the upper right panel, normalizes the $CD_5$ index by the average number of backward citations of all patents of the corresponding two-digit WIPO technology and publication year: $N_{norm}^{mean} = N_F + N_B + N_R-N_b^{mean}$. In cases where $N_R-N_b^{mean}<0$, we set $N_R-N_b^{mean}=0$ as in \cite{park2023papers}. The third variant, shown in the lower left panel, is a $CD_5$ index proposed by \cite{leydesdorff2021proposal} and considers in the nominator only forward citations made to the focal patent but not to its predecessors: $DI^*=\frac{N_F}{N_F + N_B + N_R}$. The fourth variant, shown in the lower right panel, is a $CD_5$ index proposed by \cite{bornmann2020disruption} and includes in the denominator only forward citations made solely to the focal patent and to the focal patent and parts of its predecessors: $DI_{1}^{no~k}=\frac{N_F-N_B}{N_F+N_B}$.
 \end{minipage}
\end{figure}

\clearpage

\subsection{Tables}

\begin{table}[ht]
\caption{Truncated vs. Non-Truncated Distribution of Patents' $CD_5$ Index for 1980}
\centering
\begin{tabular}{lrlrrrrr}
  \hline
   & & \multicolumn{3}{c}{Truncation}    \\ 
  No Truncation   &  -1, 0 & 0, 0.25 & 0.25, 0.5 & 0.5, 0.75 & 0.75, 1 \\ 
  \hline
 -1, 0 &  1.00 & 0.18 & 0.19 & 0.19 & 0.25 \\ 
 0, 0.25 &  0.00 & 0.82 & 0.51 & 0.34 & 0.33 \\ 
 0.25, 0.5 &  0.00 & 0.00 & 0.30 & 0.32 & 0.22 \\ 
 0.5, 0.75 &  0.00 & 0.00 & 0.00 & 0.16 & 0.05 \\ 
 0.75, 1 &  0.00 & 0.00 & 0.00 & 0.00 & 0.15 \\ 
   \hline
\end{tabular}
\label{tab:conv_matrix_trunc}
\vspace*{0.5cm}
\footnotesize
\begin{tablenotes}
\item[1] The table shows how the $CD_5$ values of patents of a particular group calculated using the \cite{park2023papers} methodology are distributed across the $CD_5$ values using the methodology without truncation. 
Reading example: The last row of the last column shows that only 15 percent of all patents with a $CD_5$ index between 0.75 and 1 and calculated by the \cite{park2023papers} methodology actually belong to the same group when all backward citations are taken into account. 
\end{tablenotes}
\end{table}

\begin{table}[ht]
\caption{Truncated vs. Non-Truncated Distribution of Patents' $CD_5$ Index for 2016}
\centering
\begin{tabular}{lrlrrrrr}
  \hline
   & & \multicolumn{3}{c}{Truncation}    \\ 
  No Truncation and Patent application   &  -1, 0 & 0, 0.25 & 0.25, 0.5 & 0.5, 0.75 & 0.75, 1 \\ 
  \hline
 -1, 0 & 0.69 & 0.26 & 0.18 & 0.18 & 0.19 \\ 
 0, 0.25 & 0.29 & 0.71 & 0.63 & 0.52 & 0.56 \\ 
 0.25, 0.5 & 0.01 & 0.02 & 0.15 & 0.18 & 0.10 \\ 
 0.5, 0.75 & 0.00 & 0.00 & 0.03 & 0.10 & 0.04 \\ 
 0.75, 1 & 0.00 & 0.00 & 0.01 & 0.02 & 0.12 \\ 
   \hline
\end{tabular}
\label{tab:conv_matrix_appl}
\begin{tablenotes}
\footnotesize
\item[1] The table shows how patents belonging to a particular group of $CD_5$ values calculated with the \cite{park2023papers} methodology are distributed across the $CD_5$ values using the methodology without truncation and considering citations to patent applications. 
Reading example: The last row of the last column shows that only 12 percent of all patents with a $CD_5$ index between 0.75 and 1 calculated by the \cite{park2023papers} methodology belong to the same group when all backward citations and citations to patent applications are taken into account. 
 \end{tablenotes}
\end{table}

\begin{table}[ht]
\caption{Aggregation of Technologies}

\centering
\footnotesize
\begin{tabularx}{\textwidth}{lll}
  \hline \hline
WIPO Field ID & WIPO Field Name & Aggregate Technology  \\ 
  \hline
1 & Electrical machinery, apparatus, energy & Electronics \\ 
  2 & Audio-visual technology & Electronics \\ 
  3 & Telecommunications & Electronics \\ 
  4 & Digital communication & IT \\ 
  5 & Basic communication processes & Electronics \\ 
  6 & Computer technology & IT \\ 
  7 & IT methods for management & IT \\ 
  8 & Semiconductors & IT \\ 
  9 & Optics & Instruments \\ 
  10 & Measurement & Instruments \\ 
  11 & Analysis of biological materials & Chemical \\ 
  12 & Control & Instruments \\ 
  13 & Medical technology & Pharma \\ 
  14 & Organic fine chemistry & Chemical \\ 
  15 & Biotechnology & Chemical \\ 
  16 & Pharmaceuticals & Pharma \\ 
  17 & Macromolecular chemistry, polymers & Chemical \\ 
  18 & Food chemistry & Chemical \\ 
  19 & Basic materials chemistry & Chemical \\ 
  20 & Materials, metallurgy & Chemical \\ 
  21 & Surface technology, coating & Chemical \\ 
  22 & Micro-structural and nano-technology & Chemical \\ 
  23 & Chemical engineering & Chemical \\ 
  24 & Environmental technology & Chemical \\ 
  25 & Handling & Mechanical \\ 
  26 & Machine tools & Mechanical \\ 
  27 & Engines, pumps, turbines & Mechanical \\ 
  28 & Textile and paper machines & Mechanical \\ 
  29 & Other special machines & Mechanical \\ 
  30 & Thermal processes and apparatus & Mechanical \\ 
  31 & Mechanical elements & Mechanical \\ 
  32 & Transport & Mechanical \\ 
  33 & Furniture, games & other \\ 
  34 & Other consumer goods & other \\ 
  35 & Civil engineering & other \\ 
   \hline
\end{tabularx}
\label{tab:tech_field}
\vspace*{0.5cm}
\begin{tablenotes}
\item[1] The WIPO technologies are defined in \cite{schmoch2008}. The USPTO assigns each patent to one or more of these technologies. The assignment can be found in the data available in PatentsView. 
 \end{tablenotes}
\end{table}

\clearpage

\bibliographystyle{elsarticle-harv}
\bibliography{main.bib}

\end{document}